# Molecular Insights into the Crystallization of 4'-Hydroxyacetophenone from Water: Solute Aggregation, Liquid-Liquid Phase Separation and Polymorph Selection

Carlos E. S. Bernardes,[a,*] Ricardo G. Simões,[a] M. Soledade C. S. Santos,[a,*] Pedro L. T. Melo,[a] Ângela F.S. Santos,[a] Stéphane Veesler,[b] Manuel E. Minas da Piedade[a,*]

[a] *Centro de Química Estrutural, Institute of Molecular Sciences, Faculdade de Ciências, Universidade de Lisboa, Campo Grande, 1749-016 Lisboa, Portugal.*

[b] *Aix-Marseille Université, CNRS, CINaM UMR 7325, 13288 Marseille, France.*

# Abstract

In this work insights into the structural rearrangements occurring in aqueous solution, prior to the nucleation of different 4'-hydroxyacetophenone (HAP) forms from water were obtained, through a combination of thermomicroscopy, micro-differential scanning calorimetry, density and speed of sound measurements, and molecular dynamics simulations. The results confirmed our previous observation that cooling crystallization of HAP is intermediated by liquid-liquid phase separation (LLPS) and, depending on the initially selected concentration range, selectively leads to the formation of different crystal forms. Analysis of the solution properties before the onset of LLPS revealed that, in the HAP mole fraction range $x_{HAP} < 0.004$ (Zone I), where hydrate H2 ultimately crystallizes, small, solvated clusters are initially present in solution, which remain approximately invariant in size, shape and HAP/$H_2O$ proportion as the temperature decreases. For the $x_{HAP} > 0.005$ range (Zone III), where anhydrous form I crystallizes, large HAP/water aggregates (that can even percolate the whole system as $x_{HAP}$ approaches the 0.005 limit) are already initially present in solution. As cooling progresses, they become more compact, a process accompanied by a reduction in water content, and which is more significant as the solution concentration increases. The $0.004 < x_{HAP} < 0.005$ (Zone II) range corresponds to a transition region where, as $x_{HAP}$ increases, the physical properties of the solution initially evolve from those typical of Zone I and, at a certain point, abruptly change and start converging to those typical of Zone III. In all zones, the colloidal particles formed upon LLPS (from which crystallization results) can also reduce their water content on cooling, but the extent of this process increases as $x_{HAP}$ moves from Zones I and II, where hydrates are formed, to Zone III, where anhydrous Form I is produced.

**Introduction**

Crystallization from solution is one of the most important processes by which solid materials are either originated in nature or synthetically produced through a variety of manufacturing and purification processes.[1] It seems now well established that the outcome of crystallization strongly depends on the early stages of the molecular self-assembly process leading to crystal formation. The microscopic description of these events is, however, quite difficult, since the identification of the sequence of aggregates that originate a particular crystal form (crystal nuclei), the time scale of their formation, and the influence of a variety of conditions that are hard to control in practice (e.g. the presence of dust, interference of wall inhomogeneities), pose considerable challenges to experimental[2-4] or computational approaches.[5,6] Nevertheless, efforts to obtain molecular level information on crystal genesis are very important, not only from a fundamental point of view (e.g. development of nucleation theories),[1,7-10] but also to provide guidelines on how to achieve better control over selective crystal form production, an aspect with considerable impact in fine chemical sectors such as pharmaceuticals (e.g., more efficient and greener processes, better products).

4'-Hydroxyacetophenone (HAP, Figure 1a), also known as *piceol*, is a compound with a wide range of applications (e.g. production of agrochemicals, flavors, fragrances, or active pharmaceutical ingredients)[11-14] that proved to be an excellent model to investigate fundamental aspects of crystallization and related phenomena. Indeed, a number of studies have shown that HAP has a rich crystal form landscape, with two enantiotropically related polymorphs (anhydrous phase cr I and cr II) and three hydrates (H1, H2, and H3) identified up to now.[15-20] It also exhibits a unique behavior when subject to cooling crystallization from water as, depending on the concentration range of the initial solution, it is possible to (*i*) selectively obtain anhydrous form I, or two different hydrates (H2 and H3) and (*ii*) observe a metastable liquid-liquid phase separation (LLPS) phenomenon mediating the formation of form I and H3 (Figure 1b).[17,21] The HAP-water combination, therefore, provides a rare opportunity to investigate the relationship between molecular organization in solution and the different pathways leading to different organic crystal forms, for a binary system where LLPS can precede crystallization.

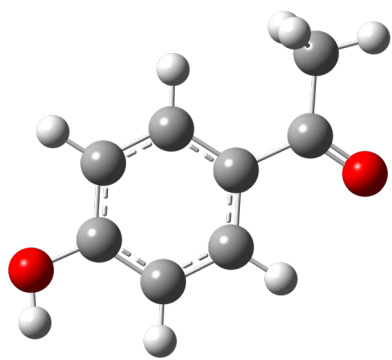 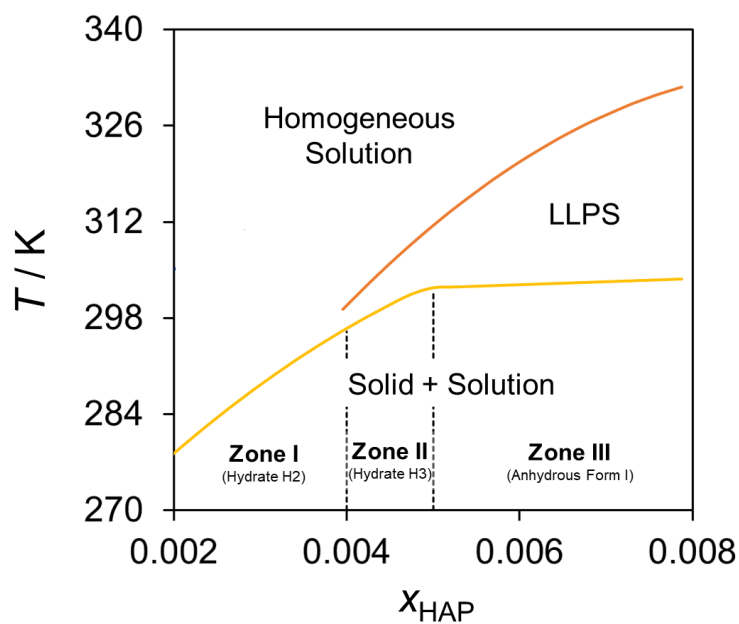

(a)                                   (b)

**Figure 1.** (a) Molecular structure of 4'-hydroxyacetophenone. (b) Cooling crystallization diagram for the 4'-hydroxyacetophenone-water system in the range $0.002 < x_{HAP} < 0.008$, obtained from previously reported experimental data corresponding to cooling crystallization runs carried out at 0.2-0.4 K min$^{-1}$.[17] The orange and yellow lines correspond to the onsets of LLPS and crystallization, respectively.

An important tool at the outset is the $T$-$x_{HAP}$ diagram ($T$ represents temperature and $x_{HAP}$ the molar fraction of 4'-hydroxyacetophenone) associated with the cooling crystallization of HAP from water in the composition range of interest, which in the present case corresponds to $0.002 < x_{HAP} < 0.008$ (Figure 1b). Previously reported data indicated that, as illustrated in Figure 1b, three $x_{HAP}$ ranges can be distinguished where different HAP phases crystallize when cooling is performed at rate in the range 0.2-0.4 K min$^{-1}$:[17] Zone I, $x_{HAP}$ < 0.004, where the crystallization of hydrate H2 is observed; Zone II, 0.004 < $x_{HAP}$ < 0.005, corresponding to the formation of hydrate H3; and Zone III, $x_{HAP}$ > 0.005, leading to the crystallization of anhydrous form I. The diagram also highlights the conditions for which LLPS was found to precede crystallization.[17] It should be noted that this is not an equilibrium diagram since, given enough time and depending on the experimental conditions, some of the phases shown evolve to more stable ones.[17] Links between the solution structure and the crystallization outcome are, however, more likely to be found if the initially detected phases are considered. Cooling crystallization runs analyzed by nuclear magnetic resonance (NMR) and dynamic light scattering (DLS) through the LLPS domain further showed that:[22] (*i*) for concentrations within Zone III, the size of the colloidal particles progressively decreases, due to the gradual release of water to the aqueous phase, until an anhydrous precursor of HAP polymorph I is formed before the onset of crystallization; (*ii*) the colloidal particles in Zone II show negligible size variation on cooling, suggesting that hydrate H3 results from the rearrangement of the solute and solvent molecules in the particle aggregates, without significant loss of solvent.

In this work, new insights into the LLPS phenomenon and molecular aggregation processes behind the crystallization of different 4'-hydroxyacetophenone (HAP) forms from water were obtained through a combination of thermomicroscopy, micro-differential scanning calorimetry ($\mu$DSC), density and speed of sound measurements, and molecular dynamics (MD) simulations.

## Materials and Methods

**Materials.** 4'-hydroxyacetophenone (HAP, Fluka) was purified by sublimation at 368 K and 13 Pa, as previously described.[15,17] The reported procedure typically leads to samples with a mass fraction >0.999. All solutions were prepared from water purified in a Milli-Q system from Millipore (resistivity, 18 MΩ cm).

**Thermomicroscopy.** The thermomicroscopy apparatus consisted of a Nikon Eclipse TE2000-U microscope equipped with an ANACRISMAT stage composed of a temperature-controlled unit, thermostated by Peltier elements, which holds a 1 cm$^3$ quartz cell containing the solution.[23] HAP-H$_2$O mixtures with $x_{HAP}$ of 0.0035, 0.042, and 0.064 were prepared and heated to 353 K to solubilize HAP. The solutions obtained were transferred to the quartz cell, and placed in the temperature-controlled unit that was also set to 353 K. After thermal stabilization for ~10 min, cooling, without stirring, at a rate of 1 K min$^{-1}$, was started, and images were simultaneously recorded every 5 s.

**Micro-Differential Scanning Calorimetry ($\mu$DSC).** $\mu$DSC experiments were carried out on a VP-DSC from Microcal. Mixtures of HAP and water corresponding to $x_{HAP}$ of 0.002, 0.004, and 0.006 were prepared in 5 cm$^3$ glass vials by weighing, using a Mettler Toledo XS205 balance (±0.01 mg precision). To ensure complete HAP solubilization, the mixtures were magnetically stirred at 353 K for 2 h, prior to use. The sample and reference cells were initially set to 353 K and filled with 0.5 cm$^3$ of HAP solution and pure water, respectively. To avoid precipitation the HAP solution was transferred into the sample cell using a pre-heated syringe. The temperature program used in the experiment consisted of two steps: (*i*) equilibration at 353 K until a stable heat flow rate was observed (40 to 60 min); (*ii*) cooling to 278.15 K at a rate of 1.0 K min$^{-1}$. At least two independent runs were performed for each concentration. The experiments were controlled using the VPViewer 2000 software, and the data treatment was performed with the Origin 7 program. The heat flow rate and temperature scales of the apparatus were periodically checked as recommended by the manufacturer, by using the built-in electric calibration system and comparing the obtained temperature of lysozyme (Sigma Aldrich, Batch 53H7145) unfolding transition, $T_{trs}$, with the value $T_{trs}$ = 331.2 K given in the IUPAC Technical Report prepared by Hinz and Schwarz.[24]

**Density and Speed of Sound Measurements.** Simultaneous density, $\rho$, and speed of sound, $u$, measurements were performed in an Anton Paar density and sound speed meter DSA-5000, with temperature control of ±0.001 K and accuracy of 0.01 K. The calibration of the apparatus was checked at 293.15 K using Milli-Q water, previously degassed for 30 min in an ultrasound bath, as recommended by the manufacturer. The values $u$ = 1482.66 m s$^{-1}$ and $\rho$ = 998.203 kg m$^{-3}$,[25,26] for speed of sound and density in pure water were considered as references.

All solutions were prepared by weight, using a Mettler Toledo XS205 balance with a precision of ±0.01 mg. The solute and solvent were mixed in a flask and sealed with a Teflon lid to prevent water evaporation. After the solution components were weighed, the flask was transferred to a water bath at 343 K and magnetically stirred until complete dissolution of the solid. The solution was then stored overnight, in an oven at 358 K, and transferred into the apparatus using a heated syringe to avoid precipitation. The measurements were performed in steps of 2 K from 343.15 K to 301.15 K, using solutions with HAP mole fractions between 0.0013 and 0.0069.

**Molecular Dynamics (MD) Simulations**. All MD simulations were performed with GROMACS.[27,28] Three HAP solutions with mole fractions of 0.002, 0.004, and 0.006, contained in a cubic box, were considered. The procedure was as follows: (*i*) an initial configuration corresponding to a random distribution of HAP and H$_2$O molecules (details are given in the Supporting Information), with a density of 0.8 g cm$^{-3}$, was generated. (*ii*) The system was equilibrated through a three-step sequence: $T$ = 360 K and $p$ = 100 bar during 1 ns; $T$ = 360 K and $p$ = 1 bar for 1 ns; $T$ = 360 K and $p$ = 1 bar during 10 ns. (*iii*) After equilibration, constant density and energy were observed and a 100 ns production stage was run at 360 K and 1 bar, during which the box configuration was recorded every 100 ps. (*iv*) The solution was cooled by 10 K, the system was re-equilibrated for 10 ns, and a new 100 ns production stage was run. (*v*) The cooling/equilibration/production sequence was repeated until a final temperature of 280 K was reached. In all simulations, the timestep was 2 fs, the temperature and pressure were controlled using a Nosé-Hoover thermostat and a Parrinello−Rahman barostat, respectively, a cutoff of 15 Å was selected, and Ewald summation corrections were used to account for electrostatic interactions beyond this limit.

The OPLS-AA[29] and TIP4P 2005[30] force field parametrizations were chosen to model HAP and $H_2O$, respectively. The Lennard-Jones (LJ) cross-interaction terms for different atomic types were computed as geometric means of the LJ parameters of each individual atom.

The GROMACS input files were prepared with DLPGEN,[31] and PACKMOL.[32] The HAP aggregation patterns and the HAP-$H_2O$ interactions were analyzed using the program AGGREGATES.[33]

## Results and Discussion

As mentioned in the Introduction, a previous study of HAP crystallization from water, using cooling rates of 0.2-0.4 K min$^{-1}$, showed that three different solid phases can first separate from the mother liquor depending on the initial concentration of the solution: hydrate H2, for $x_{HAP} < 0.004$; hydrate H3, when $0.004 < x_{HAP} < 0.005$; and the anhydrous polymorph I, for $x_{HAP} > 0.005$.[17] All these phases are metastable at ambient temperature. Indeed, if maintained in contact with the solution at 293 K, given enough time, they evolve to a third hydrate, dubbed H1 (in a matter of seconds for H3, hours for anhydrous form I, and months for H2).[17] It was also evidenced that the formation of H3 and anhydrous form I were mediated by a LLPS phenomenon (Figure 1b).[17] Finally NMR and DLS experiments[22] further suggested that the formation of polymorph I on cooling through the LLPS domain involves a progressive water loss by the colloidal droplets, while in the case of hydrate H3 hardly any water is lost and the crystals are likely to result from the rearrangement of HAP and H$_2$O molecules inside the colloidal droplets. In this work, new insights into the extent of the LLPS domain and aggregation processes that lead from a homogenous solution to the formation of HAP anhydrous phase I, H2, and H3 were obtained by combining information from thermomicroscopy, $\mu$DSC, density and speed of sound measurements, with MD simulation results.

**Thermomicroscopy.** Figure 2 shows thermomicroscopy images acquired close to the crystallization onset of HAP, on cooling from 353 K, at a rate of 1 K min$^{-1}$. Figures 2a-c refer to the formation of form I from a solution with $x_{HAP}$ = 0.0064 (Zone III in Figure 1b) and Figures 2d-f illustrate the formation of hydrate H3 from a solution with $x_{HAP}$ = 0.0042 (Zone II in Figure 1b). Movies corresponding to these processes are given as Supporting Information. When the outcome of crystallization was form I (Figure 1b, Zone III), LLPS was initially detected at 324 K (image not recorded) and persisted until 288 K (Figure 2a) where, as shown in Figures 2b-c, crystallization started. In the case of H3 formation (Figure 1b, Zone II) LLPS was noted from 299 K (image not recorded) to 283 K (Figure 2d), and then crystallization began (Figures 2e-f). The $T$-$x_{HAP}$ ranges of LLPS observed by thermomicroscopy are similar to those previously noted in experiments carried out using a 100 cm$^3$ reactor with stirring (303 K for Zone II and 324 K for Zone III).[17,34] This suggests that LLPS occurrence is, apparently, unaffected by the solution container and stirring conditions, a fact that is relevant for the direct comparison of the present results with those discussed in the next sections. The temperature onsets of crystallization detected by thermomicroscopy (288 K for form I and 283 K for H3) were, however, 15-20 K lower than those obtained in the stirred reactor experiments (~303K for form I and ~298 K for H3).[17] This indicates that a larger supersaturation level is needed for crystallization to occur when stirring is removed and/or the solution volume is reduced. Such type of observation is not unexpected and had been previously reported.[35,36]

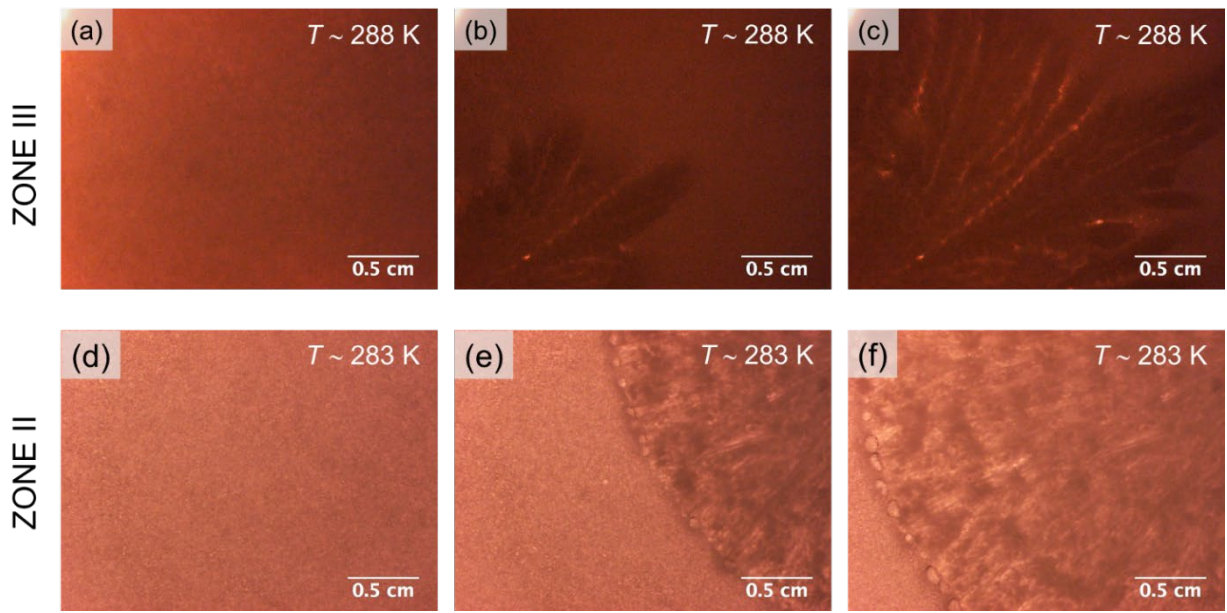

**Figure 2.** Thermomicroscopy images of the beginning crystallization of: (a)-(c) form I from a solution with $x_{HAP}$ = 0.0064 (Zone III in Figure 1); (d)-(f) hydrate H3 from a solution with $x_{HAP}$ = 0.0042 (Zone II in Figure 1). The cooling rate was 1 K min$^{-1}$.

A more complex crystallization behavior was noted when a solution with a concentration $x_{HAP} = 0.0035$ corresponding to Zone I (Figure 1b) was investigated. Previous studies in this composition range showed only the crystallization of H2, without any evidence of preceding LLPS.[17] This was, most likely, due to the impossibility of the available turbidity and imaging detection techniques to separate LLPS from an almost concomitant crystallization phenomenon. The present thermomicroscopy experiments (Figure 3) revealed that LLPS does indeed occur and is immediately followed by HAP crystallization. This is illustrated in Figure 3a, which refers to an instant close to the crystallization onset at ~295 K, and in Figure 3b, where LLPS (evidenced by a darkening of the solution) coexists with a few initially formed crystals. Another important aspect refers to the nature of the crystalized material. Initially (Figure 3c) translucid crystals are formed. After a few seconds, however, the crystals become opaque, indicating their transformation into a new phase (note the initially single-phase crystal A in Figure 3c that shows evidence of two phases in Figure 3d). As crystallization progresses it is possible to conclude that the opaque crystals correspond to the final H2 solid phase (Figure 3e-f). The nature of the translucid intermediate could not be determined.

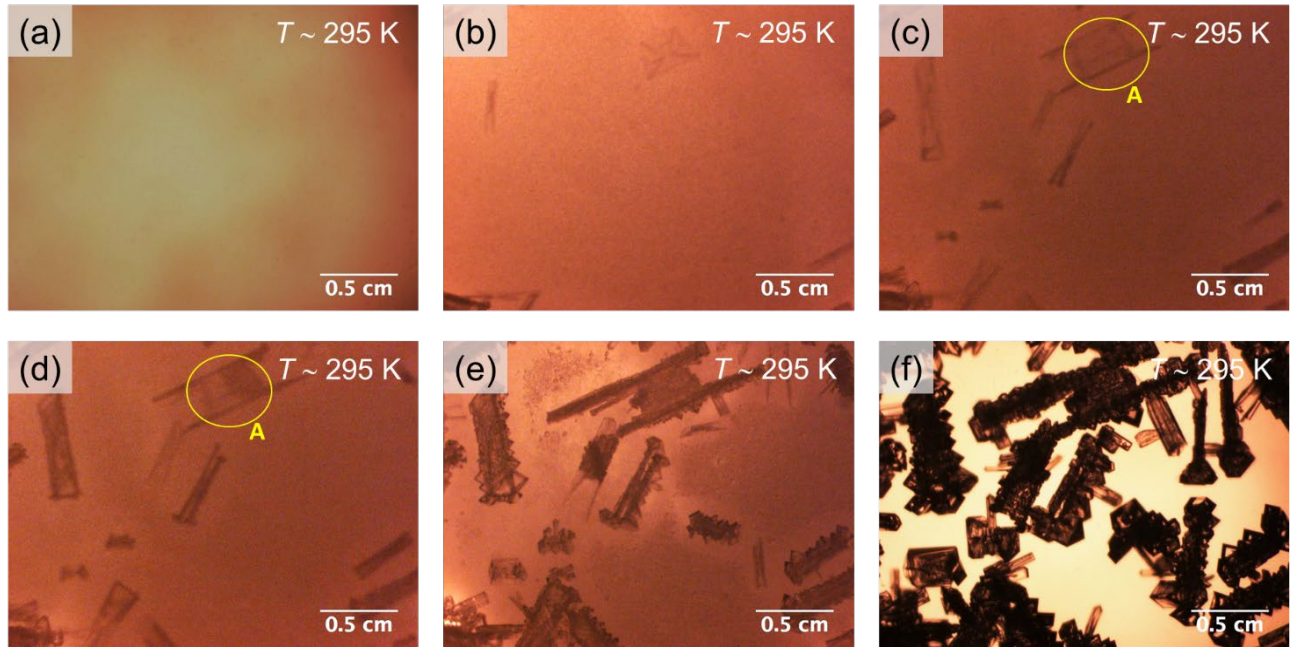

**Figure 3.** Thermomicroscopy images of main events occurring close to the crystallization onset ($T \sim 295$ K) when a HAP solution with $x_{HAP} = 0.0035$ (Figure 1b, Zone I) is cooled from 353 K, at a rate of 1 K min$^{-1}$. The yellow circle highlights a crystal (A) that is undergoing a phase transition: single phase in figure (c) and two-phases (adjacent light and dark domains) in figure (d).

**Micro-Differential Scanning Calorimetry ($\mu$DSC).** Figure 4a shows typical calorimetric curves ($\Delta\phi$ - $T$, where $\Delta\phi$ represents the heat flow difference between the sample and reference cells) recorded upon cooling HAP solutions with mole fractions $x_{HAP} = 0.002$ (Zone I in Figure 1b), $x_{HAP} = 0.004$ (Zone II in Figure 1b), and $x_{HAP} = 0.006$ (Zone III in Figure 1b), from 353 K to 278 K. The overall patterns observed in Zones II and III (blue and red curves) are significantly different from that corresponding to Zone I (green curve) and will be analyzed separately.

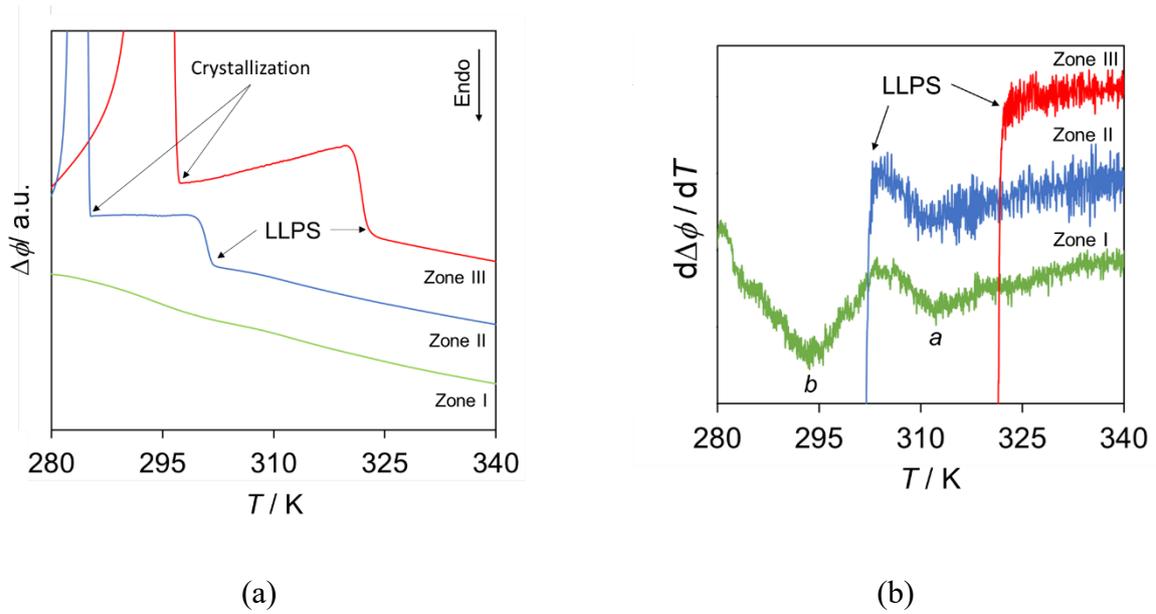

(a) (b)

**Figure 4**. (a) $\mu$DSC curves showing the evolution of the heat flow $\Delta\phi$ on cooling HAP aqueous solutions with $x_{HAP}$ = 0.002 (green; Zone I in Figure 1), $x_{HAP}$ = 0.004 (blue; zone II in Figure 1), and $x_{HAP}$ = 0.006 (red; Zone III in Figure 1). (b) First derivative of the $\Delta\phi$ - $T$ curves.

In the case of $x_{HAP}$ = 0.004 (Zone II) and $x_{HAP}$ = 0.006 (Zone III), two main exothermic events are clearly noted, which correspond to LLPS and HAP crystallization. The LLPS onsets are detected at ~323 K ($x_{HAP}$ = 0.006) and ~300 K ($x_{HAP}$ = 0.004). These values are very close to those found under analogous conditions by thermomicroscopy (324 K and 299 K, respectively, see above) and in the stirred reactor experiments (321 K and 299 K),[17] indicating that the phase separation process is not significantly affected by the solution hydrodynamics. The subsequent crystallization events are noted at 298 K (Zone III) and 285 K (Zone II). These two onset temperatures differ by 2-11 K (absolute differences) from those obtained by thermomicroscopy (288 K and 283 K for Zone III and Zone II, respectively) and stirred reactor experiments[17] (303 K and 296 K for Zones III and II, respectively). The better reproducibility and method independence of the LLPS temperature onsets compared to the crystallization onsets is not unexpected, given that the former phenomenon is less sensitive to volume and hydrodynamic effects than the latter. A further interesting feature is that, although the $\Delta\phi$- $T$ curves obtained for Zones II and III have a similar overall aspect, their heat flow patterns in the temperature range between the LLPS and crystallization onsets are different (Figure 4a): while for $x_{HAP}$ = 0.006 (Zone III) a progressive $\Delta\phi$ decrease is observed on cooling, in the case of $x_{HAP}$ = 0.004 (Zone II) $\Delta\phi$ remains approximately constant. This may reflect the previous finding by DLS and NMR[22] that, during cooling: (*i*) the colloidal particles formed in Zone III progressively decrease in size by losing water until the anhydrous form I crystallizes; (*ii*) in Zone II the particle size remains essentially constant, indicating that water loss is virtually inexistent, and the outcome of crystallization is an hydrate (H3).[22] These two different patterns may, therefore, reflect a larger dynamic rearrangement of the structure of the colloidal system in Zone III than in Zone II, as suggested by the reported DLS/NMR experiments.[22]

As mentioned above, a considerably different behavior was observed for $x_{HAP}$ = 0.002 (Zone I). In this case no significant $\Delta\phi$ change, signaling LLPS or crystallization, was noted during the cooling process (green curve in Figure 4a). Given the thermomicroscopy indication that in Zone I the two phenomena take place almost concomitantly, the obtained calorimetric result suggests that for $x_{HAP}$ = 0.002 LLPS/crystallization had not yet started when the lowest temperature attainable in the experiment (278 K) was reached. Interestingly, when the first derivative of the heat flow curve was considered (Figure 4b) the obtained result evidenced two subtle thermal events with onsets at ~312 K (point *a* in Figure 4b) and ~294 K (point *b* in Figure 4b), suggesting that the solution structure is undergoing rearrangements before the eventual onset of LLPS. The instability in the volumetric/acoustic measurements detected for $x_{HAP}$ = 0.002 at, and below, 313 K (see the discussion in the next section), suggests that these thermal events may already signal incipient LLPS. As shown in Figure 4b this is also detected in Zone II but not in Zone III.

**Volumetric and Acoustic Results.** Measurements of the density, $\rho$, and speed of sound, $u$, were used to probe how the HAP solution structure (e.g., solute aggregation, changes in the hydration sphere) varies as a function of composition and temperature. Eleven solutions with HAP concentrations covering the three zones of the cooling crystallization diagram in Figure 1b were studied ($0.001327 \leq x_{HAP} \leq 0.006816$). The measurements were performed by cooling the solutions from 343 K to the onset of LLPS. At this point, instability in the $\rho$ and $u$ measurements prevented the acquisition of additional data. These studies, therefore, refer to homogenous solutions before a well-developed LLPS occurs.

The results of the $\rho$ and $u$ measurements are illustrated in Figure 5 (data given in Tables S1 and S2 of the Supporting Information), from which the following conclusions may be extracted: (*i*) For $T \leq 337$ K, both properties show a discontinuity in the concentration range $0.004 < x_{HAP} < 0.005$ (Zone II), which is not apparent at higher temperatures. (*ii*) Outside the discontinuity range $\rho$ monotonically increases with $x_{HAP}$, with an approximately constant slope regardless of the temperature. (*iii*) The speed of sound also increases with $x_{HAP}$ but, the $u$ vs. $x_{HAP}$ lines show significant changes in slope (most notable at the lower *T* range) before and after the discontinuity, and the slope becomes smaller as the temperature increases. (*iv*) In the case of the solution with $x_{HAP} < 0.00254$ (Zone I) an instability in the measurements (continuous increase in $\rho$ and $u$ measurements; see Tables S1 and S2 in the Supporting Information) was noted below ~313 K, which approximately corresponds to the change in the *μ*DSC heat flow derivative in Figure 4b. This suggests the calorimetric and density/speed of sound observations may be related. All these features point to changes in the structure of the solution as $x_{HAP}$ and *T* vary, which must be particularly abrupt in the discontinuity region, and have a likely impact in the crystallization outcome.

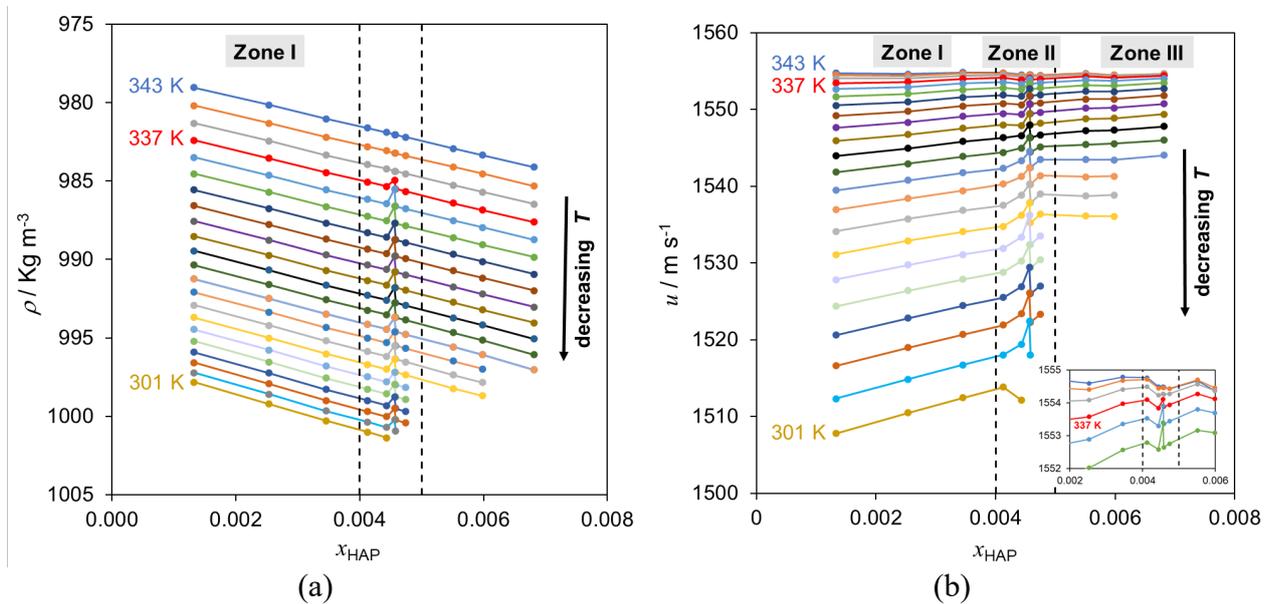

(a)          (b)

**Figure 5.** (a) Density, $\rho$, and (b) speed of sound, $u$, variation as a function of the temperature and molar fraction of the HAP solutions in water. The vertical dash lines correspond to the approximate boundaries of the different zones of the cooling crystallization diagram in Figure 1.

To obtain molecular insights into the nature of the discontinuity and, more generally, into the solute aggregation and solvation processes that, for different $x_{HAP}$ compositions, may be behind changes in solution structure along the cooling pathway preceding LLPS and crystallization, the molar volume, $V_m$, the molar isentropic compressibility, $\kappa_s$, the apparent molar volume of the solute, $V_{\phi,2}$, the relative association, *RA*, and the acoustic impedance, *Z*, were calculated from the density and speed of sound data using the relations:[37,38]

$$V_m = \frac{1}{\rho} \frac{m_1 + m_2}{n_1 + n_2} \quad (1)$$

$$\kappa_S = \frac{1}{\rho u^2} \quad (2)$$

$$V_{\phi,2} = \frac{x_1 M_1 (\rho_1^* - \rho)}{x_2 \rho \rho_1^*} + \frac{M_2}{\rho} \quad (3)$$

$$RA = \frac{\rho}{\rho_1^*} \left( \frac{u_1^*}{u} \right)^{1/3} \quad (4)$$

$$Z = u\rho \quad (5)$$

Here the subscripts 1 and 2 refer to the solvent (water) and solute (HAP), respectively; $m_i$, $M_i$, and $n_i$, represent the mass, molar mass, and the amount of substance (number of mols) of component $i$ in the solution (either solute or solvent), respectively; and $\rho_1^*$ and $u_1^*$ are the density and speed of sound of pure water, respectively, retrieved from the literature.[26] The solution molar volume mirrors the effect of the introduction of a solute molecule on the average solution volume, while the isentropic compressibility gauges hydrophobic hydration effects. Other properties such as the solute apparent volume and relative association are sensitive to solute–solvent interactions in the solution since $V_{\phi,2}$ calculation assumes that the solvent molar volume is unchanged by the addition of solute, and $RA$ compares the extent of solvent molecular association with all types of molecular association in the presence of the solute. The acoustic impedance evaluates the medium resistance to the transmission of sound waves, a property that is particularly sensitive to the system molecular order/packing.[37,38] The obtained results are given in the Supporting Information (Tables S3 to S7).

As shown in Figure 6a, in general, $V_m$ decreases during cooling, an indication of a progressive closer packing of the molecules constituting the solution. In contrast, at constant temperature, $V_m$ increases with $x_{HAP}$ reflecting the replacement of smaller and lighter $H_2O$ by larger HAP molecules. Regardless of temperature, the slope of this increase is approximately identical throughout the whole concentration domain studied, except in a narrow concentration range of Zone II ($0.00452 < x_{HAP} < 0.00465$), where the discontinuity noted for $\rho$ and $u$ in the temperature interval 303-337 K is also apparent (Figure 6a). As shown in Figure 6b, the slope of the $V_m$ vs $x_{HAP}$ plots is positive outside the [$x_{HAP} = 0.00452$-$0.00465$; $T = 303$-$337$ K] domain, and negative inside. The $V_m$ contraction that occurs within this domain is likely to signal the formation of HAP aggregates that lead to a $H_2O$ cage breakage and to a closer packing of the molecules, because the volume shrinkage.

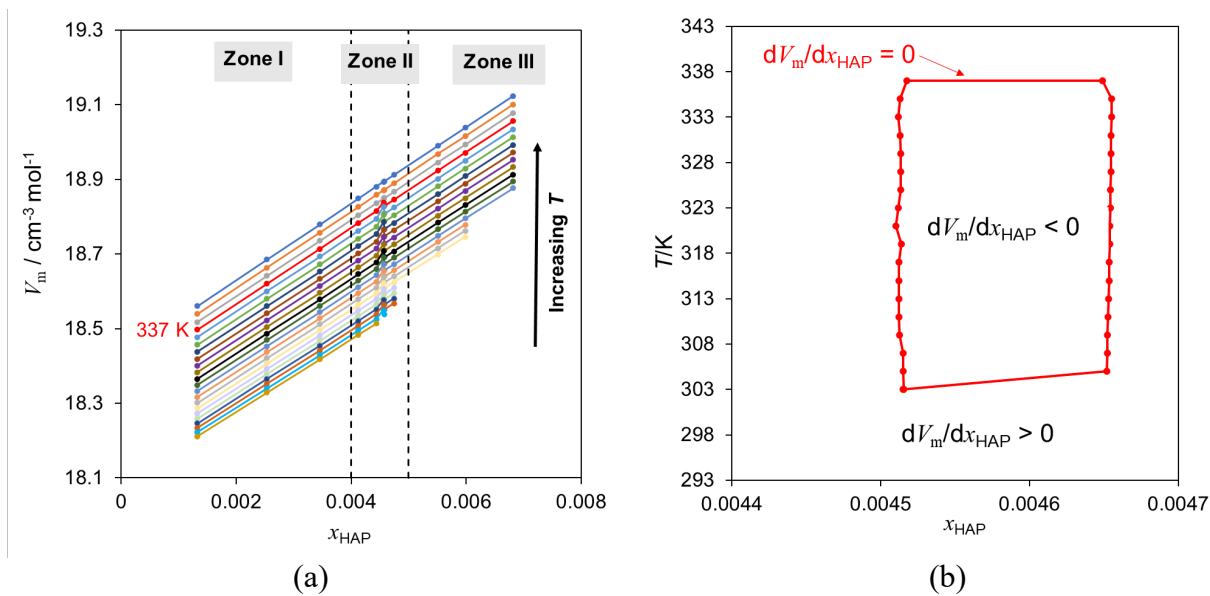

**Figure 6**. (a) Dependence of the solution molar volume, $V_m$, on temperature and HAP molar fraction, $x_{HAP}$, in water. The vertical dash lines give the approximate boundaries of the cooling crystallization diagram zones where different species crystallize from solution. (b) Boundary of the $dV_m/dx_{HAP}$ sign change (red line).

associated with the collapse of the water 3D hydrogen bond network is not compensated by the formation of HAP aggregates. These observations suggest that in Zone I the water 3D hydrogen bond (H-bond) network can elastically respond to the progressive insertion of HAP molecules, which are isolated or forming small aggregates, without any major disruption. When Zone II is reached, the aggregates have grown to a size that can no longer be accommodated by water without a major structural rearrangement. After the discontinuity, the initial (Zone I) $V_m$ vs $x_{HAP}$ slope is regained indicating that the new solution structure is now able to progressively accommodate the larger growing aggregates, throughout Zone III, without an abrupt disruption.

The isentropic compressibility is generally considered to be more sensitive to hydration changes than $V_m$.[39] The dependency of $\kappa_s$ on $T$ and $x_{HAP}$ is illustrated in Figure 7a. The following main conclusions can be drawn from these data. At all temperatures: (*i*) $\kappa_s$ decreases with the increase of $x_{HAP}$, except where the above-mentioned disruption is observed within Zone II. (*ii*) The negative and constant slopes of the $\kappa_s$ vs. $x_{HAP}$ plots in Zones I and III indicate that as $x_{HAP}$ increases the solution becomes less compressible and the structure progressively evolves without abrupt rearrangements. (*iii*) The larger negative slope in Zone I compared to Zone III indicates that the solution structure is more sensitive to HAP addition in the former than in the latter zone. Such behavior may reflect the destruction of a larger number of water-water H-bonds per HAP molecule added to the solution, in Zone I than in Zone III. This interpretation is compatible with the view conveyed by the $V_m$ analysis and further allows to distinguish Zones I and III. The larger negative $\kappa_s$ vs. $x_{HAP}$ slope in Zone I compared to Zone III suggests the formation of a larger number of HAP-$H_2O$ interactions per added HAP molecule in the former than in the latter zone. Indeed, the predominance of HAP molecules individually solvated or forming small clusters, in Zone I would necessarily translate into the formation of more HAP-H$_2$O interactions per added HAP molecule when compared to Zone III, where large solute aggregates are present and grow as $x_{HAP}$ increases (this idea is reinforced by the MD results discussed in the next section).

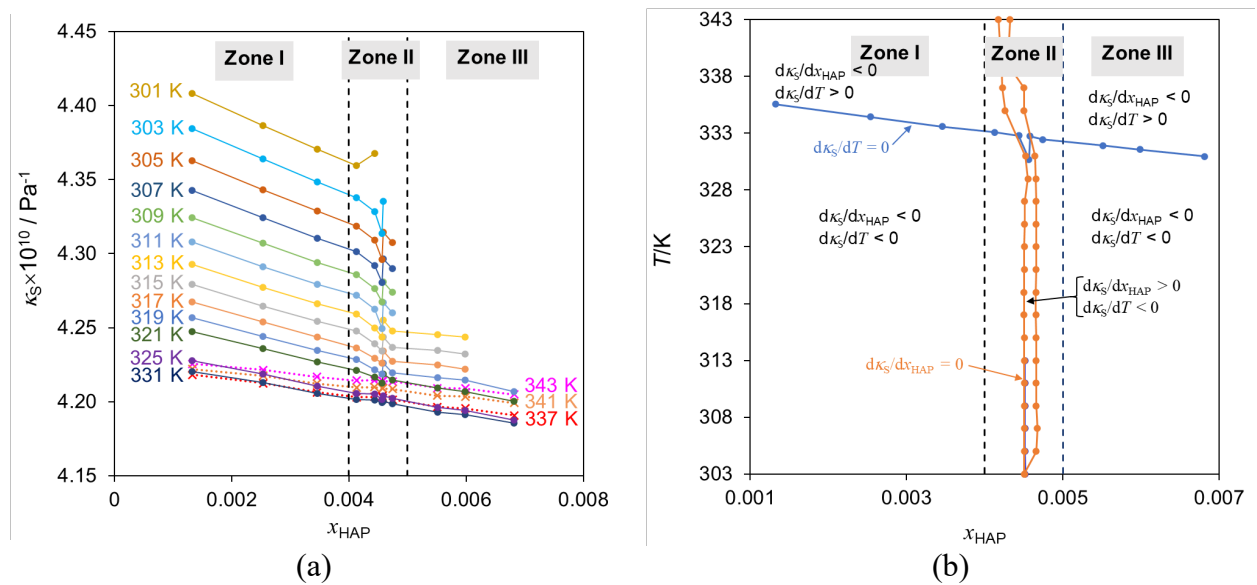

(a) (b)

**Figure 7**. (a) Dependence of solution isentropic compressibility on temperature and molar fraction of HAP in water. For the sake of clarity data at some selected temperatures were omitted and are given as Supporting Information. Solid lines and filled markers refer to measurements where $\kappa_s$ decreases with temperature. Dotted lines and cross symbols indicate measurements where $\kappa_s$ increases with temperature. The vertical dash lines give the approximate boundaries of the cooling crystallization diagram where different species crystallize from solution. (b) Boundary of the $d\kappa_s/dx_{HAP}$ (orange line) and $d\kappa_s/dT$ (blue line) sign change.

A minimum in $\kappa_s$, associated with the discontinuity (d$\kappa_s$/d$x_{HAP}$ = 0) is evidenced in Figure 7b. This singularity is consistent with a sudden increase of HAP aggregation, involving structures which are more compact and less compressible than water. Furthermore, minima in $\kappa_s$ are usually attributed to the hydrophobic effect, which, in this case may be responsible for triggering HAP aggregation to minimize water-aromatic ring interactions.[40] Indeed, recalling Uricks' equation for dispersed systems ($\kappa_s = \kappa_w \phi_w + \kappa_o \phi_o$),[41,42] that should be valid under the present experimental conditions (dilute solutions, similar liquid phase densities, and $\nu$ = 3 MHz),[43] the measured isentropic compressibility will be the volumetric weighted average of the aqueous ($\kappa_w$) and water immiscible "oil"·($\kappa_o$) phases compressibility ($\phi_w$ and $\phi_o$ represent the volume fractions of aqueous and water immiscible phases, respectively). If LLPS occurs, a $x_{HAP}$ increase leads to an $\phi_o$ increase, which will be reflected by a $\kappa_s$ decrease whenever large HAP-HAP clusters are present. Thus, even if incipient LLPS occurs, the observed behavior suggests that the aqueous phase mainly includes hydrated species (HAP molecules that are isolated or forming small aggregates) which evolve with concentration to larger and less compressible HAP aggregates that ultimately originate the denser "oil" phase droplets. Indeed, the average slope d$\kappa_s$/d$x_{HAP}$ is more negative in Zone II than in Zone III where the presence of even larger HAP-HAP agglomerates has also been hypothesized, certainly because in Zone III the addition of HAP molecules delivers a much smaller cluster volume change or, in other words, a smaller increase of $\phi_o$. On cooling, $\kappa_s$ first decreases from 343 K to 335 K and subsequently increases from 335 K to 301 K (Figure 7a). The temperature at which the inversion occurs is not exactly the same for the different concentrations studied, exhibiting a slight decrease tendency as $x_{HAP}$ increases. This is apparent in Figure 7b, where the blue line marking the $d\kappa_s/dT = 0$ frontier has a negative slope. This type of anomaly has also been reported for pure water,[25,26] where the $\kappa_s$ minimum is observed at ~334 K. Its cause has been qualitatively assigned[25,26] to an unspecified structural variation resulting from the balance between two opposing effects: a larger extent of $H_2O$ cage breakage as the temperature increases, that produces more compact structures and leads to a speed of sound increase; and the corresponding volume expansion, which contributes to a decrease in ultrasonic velocity. Because the first effect dominates at lower temperatures and the second at higher temperatures, $\kappa_s$ goes through a minimum. The same qualitative arguments have been used to explain the sound velocity minima observed in other aqueous systems, such as water-butoxyethanol in the 328-340 K range.[44] The present results, therefore, suggest that, in the concentration range probed, this phenomenon is not significantly perturbed by the presence of HAP.

Figure 7b also reveals that above 337 K the discontinuity observed in Zone II vanishes and the slope of the $\kappa_s$ vs. $x_{HAP}$ line becomes essentially constant throughout Zones I to III. This indicates that above this temperature the structure of the solution is less affected by concentration changes, a feature compatible with a dominant presence of hydrated HAP molecules, which are isolated from each other or forming loose clusters. The discontinuities found below this temperature suggest that effective solute aggregation can already be occurring in solution well above the observed onsets of LLPS or crystallization (Figure 1). This is consistent with experimental findings evidencing that self-assembled mesoscale structures can spontaneously form in unsaturated bi- or multicomponent liquid mixtures,[40,45,46] an aspect that is further supported by the MD simulations discussed below.

The results obtained for the solute apparent molar volume, $V_{\phi,2}$, show that this property decreases with decreasing temperature (Figure 8a). At a fixed temperature, in Zone I, small $V_{\phi,2}$ oscillations are noted as $x_{HAP}$ increases, signaling structural rearrangements associated with the incorporation of HAP in the water structure. Analogously to what was observed for $V_m$ and $\kappa_s$, the most notable rearrangement occurs in Zone II where, below ~337 K, a discontinuity in $V_{\phi,2}$ is observed for $x_{HAP} \sim 0.0045$. At this point the preceding slight tendency for the $V_{\phi,2}$ increase with $x_{HAP}$ is interrupted and an abrupt $V_{\phi,2}$ decrease is observed. After this decrease, $V_{\phi,2}$ remains essentially constant while entering Zone II, a tendency that extends throughout Zone III. However, as the temperature is reduced, $V_{\phi,2}$ starts to decrease as LLPS is approached, suggesting that the solvated HAP molecules aggregate and simultaneously lose hydration water, translating into a decrease in the average volume assigned to HAP. These results are consistent with the MD simulation results discussed below.

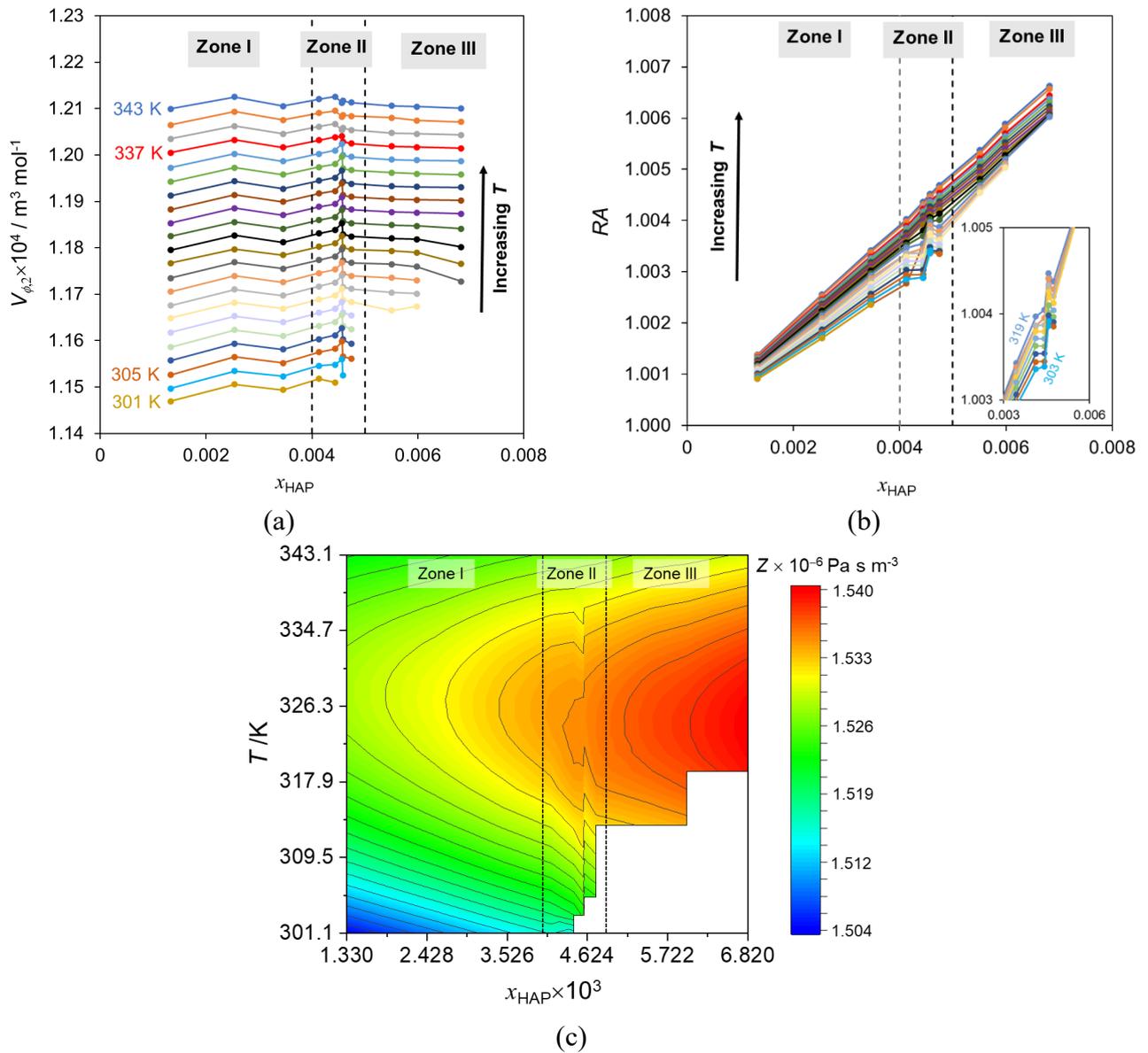

**Figure 8**. (a) Apparent molar volume of the solute, (b) relative association, and (c) acoustic impedance of the solutions of HAP in water with varying concentrations and temperatures. The vertical dash lines give the approximate boundaries of the different zones of the cooling crystallization diagram.

The results of the relative association, $RA$, are given in Figure 8b. This parameter reflects the extent of solvent association in the solution when compared to the pure solvent. As such, the destruction of the solvent structure will lead to $RA < 1$, while new solute-solvent interactions will imply $RA > 1$.[37] As can be observed in Figure 8b, $RA > 1$ for all temperatures and concentrations, indicating a larger molecular association involving HAP-HAP, HAP-W (W = water), and W-W interactions in the HAP solutions than in pure water. As for the above discussed parameters, abnormal variations in $RA$ can be observed in the concentration range $0.004 < x_{HAP} < 0.005$ for $T <$ 337K. A clear meaning could not be attached to these variations, though it is evident that the increase in $RA$ with $x_{HAP}$ is smaller in Zone II than in Zones I and III. This indicates that on average less HAP-W interactions are formed in Zone II for the same $x_{HAP}$ increase, consistent with the onset of HAP aggregation in solution at constant temperature. At ~337 K, a shift can be observed in $RA$, accompanying the shift in the isentropic compressibility with increasing HAP composition, a singularity that may be linked to the increased immobilization of the hydration shell molecules at lower temperatures.[47]

The acoustic impedance is a measure of the resistance of the medium to the passage of sound and is sensitive to the formation of new phases with different impedance. The results obtained are illustrated in Figure 8c. This figure shows a maximum in the acoustic impedance centered at the concentration range of Zone III, which extends to Zone II. Such observation suggests the solution structure is continuously evolving until the LLPS onset is observed ultimately leading to the formation of two distinct macroscopic liquid phases. The process is more evident in Zone III where the LLPS temperature domain is wider. Furthermore, at a given temperature, the maximum values of the acoustic impedance vary in the order Zone III > Zone II > Zone I, despite the irregularities noted in Zone II for $T \leq 337$ K. The sudden decrease in the acoustic impedance in Zone II may signal the release of water molecules from the HAP-W aggregates, accompanied by a decrease in aggregate size, which is more evident as $T$ decreases. This view is supported by the MD simulation results presented in the next section.

The linear behavior observed for the variation of $V_\mathrm{m}$ and $\kappa_\mathrm{s}$ with $x_\mathrm{HAP}$ in Zone I (Figures 6a and 7a) is expected for a solution of predominantly unassociated solute molecules or including small aggregates surrounded by their solvation spheres (a hypothesis corroborated by the MD data discussed in the next section). Under these conditions, acoustic methods can be employed to estimate the hydration number, $n_\mathrm{H}$,[48,49] i.e. the number of water molecules directly bound by H-bonds to each solute molecule, from the variation of the compressibility of the HAP solution compared with that of pure water.

Passynski proposed an acoustic method to calculate hydration numbers,[48] assuming that the compressibility of water molecules involved in solute solvation was negligible. Glinski and Burakowski [49] improved this methodology allowing for a non-zero hydrate cluster compressibility using the equation:

$$n_\mathrm{H} = \frac{n_\mathrm{W}\left(\kappa_S - \kappa_{S,\mathrm{W}}\right)}{n_\mathrm{HAP}\left(\kappa_{S,\mathrm{hydrate}} - \kappa_{S,\mathrm{W}}\right)} \tag{6}$$

where $n_\mathrm{w}$ and $n_\mathrm{HAP}$, are the number of moles of water and HAP in the solution, respectively, $\kappa_S$ is the solution isentropic compressibility, $\kappa_{S,\mathrm{W}}$ is the isentropic compressibility of pure water, and $\kappa_{S,\mathrm{hydrate}}$ is the isentropic compressibility of the solute in the solvation sphere. As the compressibility of the hydration water is unknown, several values ranging from zero (as initially proposed by Passynski) to $1.6 \times 10^{-10}$ Pa$^{-1}$ have been considered.[50,51] The dependence of the hydration number on temperature is displayed for the lowest studied concentration, $x_{HAP}$ = 0.001327, where no liquid-liquid separation was identified (Figure S3). The hydration number, $n_H$, increases with decreasing temperature, reflecting the closer packing in solution at lower temperatures. The value of $n_H$ varies from ~2 molecules of water bound to each HAP molecule at higher temperatures to 4-5 molecules of water at lower temperatures. It should also be noted that these values set an upper limit for the stoichiometry of the hydrated solids that may be isolated from these solutions since these results mirror the number of water molecules exhibiting a physico-chemical behavior distinct from bulk water. This is consistent with the fact that the stoichiometry of the H2 hydrate obtained in Zone I is HAP·3H$_2$O.[17]

In summary, the study of the volumetric and acoustic properties of the solutions reveals: (*i*) the existence of different solution structures in Zones I, II and III; (*ii*) Zone II corresponds to a transition region where an abrupt solution structure reorganizations occur; (*iii*) for temperatures above ~337 K homogeneous solution behavior is observed in all concentration Zones. (*iv*) Upon cooling, however, the solution structures evolve through different paths, depending on concentration, to generate various types of HAP-Water molecular aggregates that precede LLPS: in Zone I HAP-HAP aggregates that may comprise some water molecules in Zone II; and solvated aggregates in Zone III. These ideas will be further developed in the next section based on molecular dynamics simulation results.

**Molecular Dynamics Simulations.** MD simulations were used to provide microscopic level insights into the structural changes occurring in solution as temperature and concentration change, prior to LLPS development and crystallization. Three HAP concentrations, corresponding to the three zones of the cooling crystallization diagram discussed in this work, were investigated: $x_{HAP}$ = 0.002 (Zone I), 0.004 (Zone II), and 0.006 (Zone III). For each one of them, simulations were initiated at 360 K, where a homogeneous solution exists, and the temperature was subsequently decreased to 280 K in steps of 10 K. Due to computational limitations, cubic boxes with ~117 Å side length were used. Under these conditions, the liquid-liquid phase separation phenomenon experimentally detected in the $T$-$x_{HAP}$ range covered by the simulations cannot be observed, because the diameter of the emulsion droplets is much larger than the size of the box (previous experimental determinations gave droplet diameters in the range 1000 Å to 8000 Å).[22] The MD results can, nevertheless, shine light on the structural changes occurring in solution during the cooling process, that are likely to underly the observations of the volumetric and acoustic experiments discussed in the previous section, as well as the differences in the crystallization outcome observed from Zones I to III.

*Solute Aggregation*. To investigate the HAP aggregation patterns as $T$ and $x_{HAP}$ vary, and the role played by water (W) in the process, the MD trajectories were analyzed for the presence of HAP aggregates and their $H_2O$ content. For the identification of HAP aggregates, all solute molecules that contact with each other were determined, their connectivity assigned, and the probability of finding an entity with a given number of HAP molecules, $N_{HAP}$, was calculated.[33] Two HAP molecules were considered to be in contact if the distance between any of their atoms was smaller than the sum of the corresponding van der Walls radii, $r_{vdw}$,[52] plus a constant value of 0.5 Å.[33] The results for three representative temperatures are illustrated in Figure 9, from which the following conclusions can be inferred: At 360 K, independently of the concentration, most of the HAP molecules are either isolated from each other or forming relatively small aggregates. For $x_{HAP}$ = 0.002 (Zone I) these clusters have < 13 molecules (Figure 9a). Cooling the mixture leads to a small increase in the probability of finding larger HAP clusters, but they contain only less than ca. 25 molecules (less 25 % of the total number of solute molecules in the simulation box) and the probability of their formation is quite small (ca. 0.01). This contrasts with the results obtained for concentrations within Zones II ($x_{HAP}$ = 0.004) and III ($x_{HAP}$ = 0.006). In the latter cases, decreasing the temperature has a dramatic effect in the structure of the solution, since the probability of finding aggregates containing almost all HAP molecules present inside the simulation box is quite large for temperatures below ~350 K (Figure S4 in the Supporting Information). These structural differences are consistent with the results of the volumetric and acoustic measurements discussed in the previous section, showing that properties such as $u$ and $\kappa_s$ have clearly different $T$ and $x_{HAP}$ dependencies in zones I, II and III.

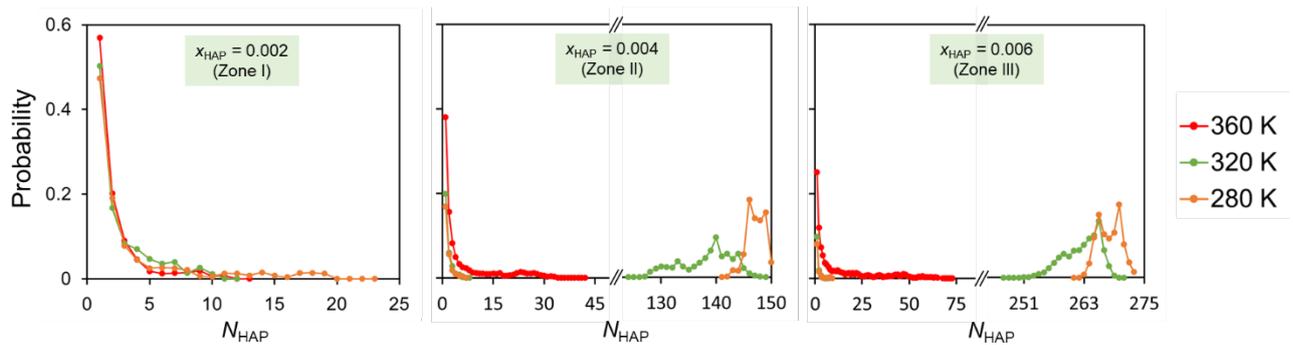

**Figure 9**. Probability of finding aggregates containing $N_{HAP}$ molecules of HAP, at different temperatures and compositions: (a) $x_{HAP}$ = 0.002, (b) $x_{HAP}$ = 0.004, and (c) $x_{HAP}$ = 0.006. The simulation boxes contained 100, 200, and 300 molecules of solute for $x_{HAP}$ = 0.002, $x_{HAP}$ = 0.004, and $x_{HAP}$ = 0.006, respectively.

*Solute-Water Interaction*. To investigate the role of water in the aggregation process as cooling progresses towards LLPS and crystallization, the $H_2O$ content of the HAP clusters was analyzed. It may be pointed out at the outset that this analysis suggested that, in line with previous experimental observations,[22] the most significant differences between Zones II and III in terms of molecular aggregation and crystallization outcome (hydrate H3 in Zone II and anhydrous form I in Zone III, Figure 1), were related to the extent of water incorporation in the HAP aggregates and how this extent decreased along the cooling process. The adopted methodology was as follows: (*i*) Initially, all clusters including HAP and water were identified by searching for 3D HAP-W networks. It was again considered that two molecules were in contact if the distance between any of their atoms was smaller than $r_{vdw}$ + 0.5 Å. (*ii*) To estimate the number of solvent molecules effectively inside the clusters, all water molecules contacting only with one HAP molecule were discarded. This minimized the effect of counting water molecules that are only present at the "cluster interface" with the solution. (*iii*) The size of each HAP-W aggregate was then determined as the total number, $N$, of HAP and W molecules it contains. (*iv*) Finally, for each $N$-size cluster, the average percentage of HAP molecules in the structure (%HAP) was computed.

The results obtained from the previously described analysis are illustrated in Figure 10, which gives %HAP at different concentrations and temperatures as a function of $N$. Figure 10 shows that: (*i*) Consistent with the view conveyed by Figure 9, larger aggregates are observed in solution on increasing the concentration and lowering the temperature. (*ii*) However, for $x_{HAP} = 0.002$ (Figure 10a), independently of the concentration and temperature, the size of the larger aggregates is still small (less than ~70 molecules). Moreover, their water content increases significantly with size and becomes approximately constant for $N > 15$ (e.g. %HAP = 50 for $N = 4$ and %HAP ~ 30 for $15 < N < 70$). This suggests that consistent with the above discussion of solute aggregation, in Zone I, HAP molecules remain essentially scattered and solvated by water along the cooling process. (*iii*) In contrast, for the higher concentrations, $x_{HAP} = 0.004$ and $x_{HAP} = 0.006$, a considerable change in aggregate size with temperature is noted. At 360 K, the total number of HAP and H$_2$O molecules in the aggregates hardly exceeds $N = 200$, with %HAP ~ 27-40. As the temperature decreases, not only larger aggregates containing most solute molecules in the simulation box emerge (as already concluded from Figure 9), but the HAP to water proportion also increases, with %HAP ~35-46 (Figures 10b,c). The comparison of %HAP in these large entities at high and low temperatures, therefore, suggests that on cooling, water molecules are progressively released to the solution bulk.

(*iv*) Moreover, the temperature dependency of the ratio between the number of water and HAP molecules in the aggregates ($N_w/N_{HAP}$) shown in Figure 11, suggests that the extent of this release is more pronounced in Zone III than in Zone II. Indeed, albeit at 360 K the aggregates of larger sizes have fewer HAP molecules for $x_{HAP}$ = 0.006 ($N_w/N_{HAP}$ = 2.2) than for $x_{HAP}$ = 0.004 ($N_w/N_{HAP}$ = 2.0), the opposite order is quickly attained as the temperature decreases (e.g. at 280 K $N_w/N_{HAP}$ = 1.43 for $x_{HAP}$ = 0.006 and $N_w/N_{HAP}$ = 1.58 for $x_{HAP}$ = 0.004). This is consistent with our previous experimental findings from $^1$H NMR and dynamic light scattering experiments that: (*i*) the dense phase droplets resulting from LLPS and from which crystallization is likely to start, have a considerably higher water content in Zone II than in Zone III; (*ii*) the $N_w/N_{HAP}$ ratio in the dense phase droplets progressively decreases on cooling, and the decrease is faster in Zone III than in Zone II; (*iii*) before crystallization the loss of solvent is complete in Zone III, where the anhydrous HAP polymorph I is produced, and incomplete in Zone II where hydrate H3 is formed.[22]

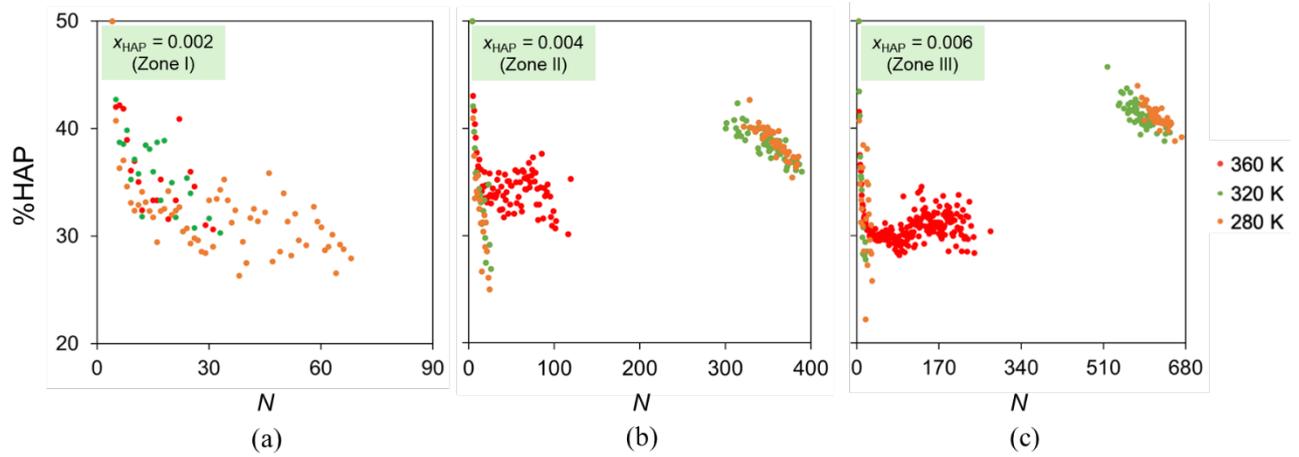

**Figure 10**. Percentage of HAP molecules, %HAP, as a function of the total number, $N$, of molecules (HAP and water) in the clusters, at different temperatures and (a) $x_{HAP}$ = 0.002, (b) $x_{HAP}$ = 0.004, and (c) $x_{HAP}$ = 0.006.

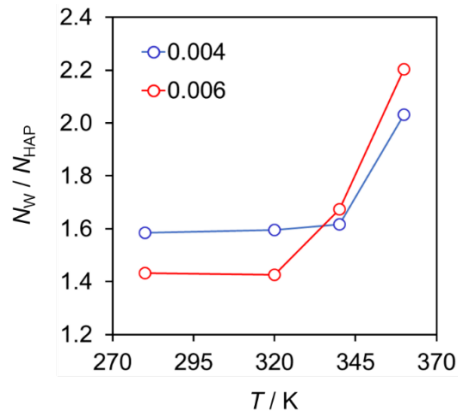

**Figure 11**. Variation of water/HAP molar ratio, $N_W/N_{HAP}$, as a function of temperature, $T$, obtained for the larger aggregates (containing more than c.a. 70 molecules), in the two most concentrated solutions, observed during the MD simulation trajectories.

It is also interesting to note that this behavior is consistent with the observed changes in properties like $\kappa_s$ (Figure 7a), $V_{\phi,2}$ (Figure 8a), and *RA* (Figure 8b), which show a larger dependence on concentration in Zone III than in Zone II. It is therefore likely that the experimental observations from volumetric and acoustic measurements are related to the change in the $N_W/N_{HAP}$ ratio of the pre-LLPS aggregates present in solution.

*Analysis of Aggregates Size and Shape.* Further insights into the structural changes experienced by solutions in Zones I-III upon cooling, were provided by analyzing the shape of the HAP-W aggregates in Figure 10 as a function of temperature and concentration. The analysis relied on the calculation of two quantities: (*i*) the largest distance, *d*, between two atoms belonging to the same aggregate (independently of the molecule) and (*ii*) the apparent volume of the aggregate, $V_{agg}$. The latter quantity was obtained using a previously described procedure,[33] that uses a series of grid points around the aggregate to define a surface, which approximately captures its shape, and allows the subsequent calculation of $V_{agg}$. Figure 12 shows $V_{agg}$ vs. *d* plots corresponding to aggregates with *N* > 70 and to the different $x_{HAP}$ zones investigated. The limit *N* > 70 was considered to discard HAP-W aggregates which essentially correspond to solvated HAP molecules. The data used to build the plots represent averages of the individual $V_{agg}$ and *d* values obtained for a given *N* size aggregate present in all MD simulation frames recorded during the production stage. Also shown in Figure 12 are $V_{agg}$ vs. $d$ lines computed for spherical (solid line), or prolate (dash lines) aggregate shapes as $V_{agg} = (4/3)\pi a^2 c$, where $a$ is the equatorial radius, and $c = d/2$ is the polar radius. The $a$ to $c$ relationships (aspect ratios) that best adapted to the MD results are indicated below and also in Figure 12. Comparison of those lines with the MD data indicated that prolate shapes are favored, because attempts to fit the results to oblate contours led to aggregates with a distance between the poles < 2 Å, which seems physically impossible.

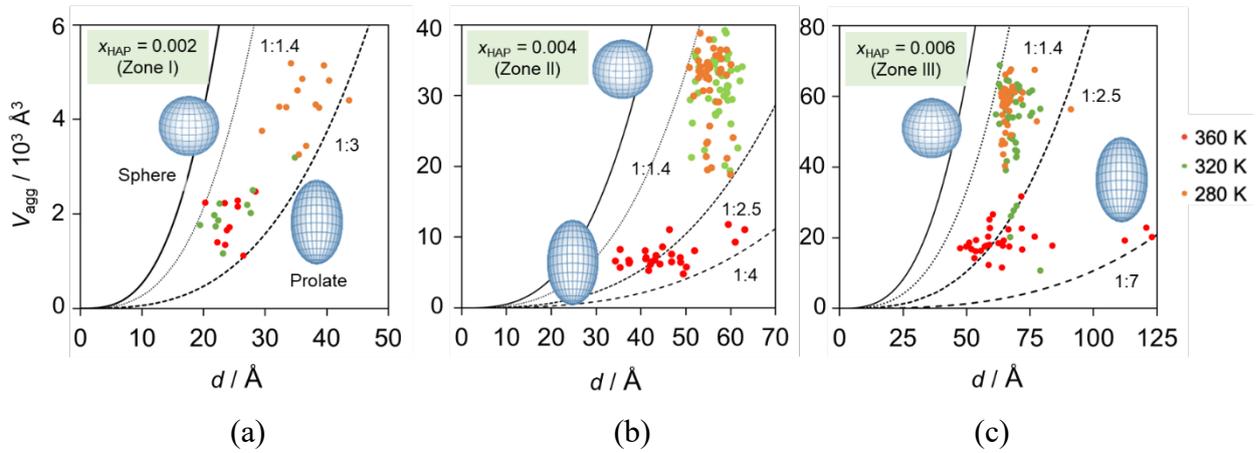

(a)                       (b)                      (c)

**Figure 12**. Apparent volume of the HAP-W aggregates, $V_{agg}$, as a function of the maximum distance between two atoms, $d$, belonging to the same aggregate given by the MD simulations, for different temperatures and HAP concentrations: (a) $x_{HAP} = 0.002$, (b) $x_{HAP} = 0.004$, and (c) $x_{HAP} = 0.006$. The lines represent the $d$ vs. $V_{agg}$ plots expected for spherical (solid line) or prolate (dash lines) aggregate shapes.

The results in Figure 12 suggest that: (*i*) when the temperature decreases, in Zone I, despite the slight increase of the number of molecules in the aggregates, their shape distribution is largely unaffected, remaining prolate with an axial ratio within 1:1.4 and 1:3. (*ii*) In the concentration ranges of Zones II and III, at 360 K, a larger dispersion of prolate shapes compared to Zone I is noted which is more pronounced in Zone III (axial ratios in the range 1:1.4 to 1:4 for $x_{HAP} = 0.004$ and 1:1.4 to 1:7 for $x_{HAP}$ = 0.006). (*iii*) Cooling the solution considerably narrows the aspect ratio range, which becomes essentially comprised between 1:1.4 and 1:2.5, both for $x_{HAP}$ = 0.004 and $x_{HAP}$ = 0.006. This lowering of the aspect ratio limits suggests that in Zones II and III the loss of water during cooling (signaled by the $N_w/N_{HAP}$ decrease and increase of %HAP), is accompanied by the formation of more spherical-like particles. (*iv*) It is also necessary to reach lower temperatures in the case of the solution with $x_{HAP}$ = 0.006, to observe a similar aspect ratio distribution as that found for $x_{HAP}$ = 0.004. These conclusions therefore suggest that the development of LLPS involves a larger rearrangement of the molecules in Zones III than in Zone II.

Thus, overall, the MD data in Figures 9-12 indicate that at higher temperatures (e.g. 360 K), the HAP-W aggregates adopt a prolate shape which becomes more elongated as $x_{HAP}$ moves from Zone I to Zone II and Zone III. For $x_{HAP}$ = 0.002 (Zone I), the HAP molecules are organized in small, scattered clusters ($N$ < 70), which do not significantly vary in size and shape upon cooling. For $x_{HAP}$ ≥ 0.004 (Zones II and III) larger ($N$ < 200) aggregates compared to $x_{HAP}$ = 0.002 are already present in solution at 360 K. These structures are very elongated and at $x_{HAP}$ = 0.006 can reach almost 125 Å in length (Figure 12c), meaning that they can percolate the entire system (the length of the cubic simulation boxes is ~117 Å). But most notable, the number of molecules in the aggregates increase (Figure 10), even though their size ($d$ in Figure 12) decreases because more spherically shaped structures are formed. The formation of these smaller spherical structures containing almost all solute molecules in the simulation box implies that a folding of the initial elongated aggregates, which promotes the loss of water, must occur during cooling. This is compatible with a reduction of surface energy, indispensable in spontaneous emulsification processes. The main difference between Zones II and III is that larger aggregates are initially found in Zone III than in Zone II, and, albeit on cooling they converge to approximately the same shape ratio, the water loss is larger and faster in Zone III than in Zone II.

These findings support the view conveyed by our previous NMR and DLS experiments[22] that, upon cooling, different evolution pathways are followed by the HAP-W aggregates initially present in solution, which essentially depend on the concentration range, and lead to different crystallization outcomes (hydrate H2 in Zone I, hydrate H3 in Zone II, and anhydrous form I in Zone III).

## Conclusions

Building on our previous findings from stirred reactor,[17] [1]H NMR, and DLS[22] experiments, new insights into the molecular processes behind the crystallization of different 4'-hydroxyacetophenone forms from water in the range $0.002 < x_{HAP} < 0.008$, were obtained by combining thermomicroscopy, micro-differential scanning calorimetry, density and speed of sound measurements, with MD simulations. The pathways that lead to hydrate H2 for $x_{HAP} < 0.004$ (Zone I), hydrate H3 for $0.004 < x_{HAP} < 0.005$ (Zone II) and anhydrous form I for $x_{HAP} > 0.005$ (Zone III) suggested by the overall results are illustrated in Figure 13. This figure tentatively condenses the following main aspects:

- **Zone I ($x_{HAP} < 0.004$):** At high temperatures, HAP molecules are predominantly isolated or form small aggregates in solution as evidenced by solution properties (volumetric, acoustic and derived quantities). Cooling does not significantly alter this arrangement until the onset of LLPS/crystallization, except for a slight increase in the amount of HAP within the particles. Based on previous proposals by Gebauer *et. al*,[53] for the crystallization of inorganic materials, it is likely that these small molecular aggregates coalesce generating LLPS, which, in this composition range, immediately rearranges to crystallize as a hydrated solid material.

- **Zone II ($0.004 < x_{HAP} < 0.005$):** This zone can be considered a transition region in the cooling crystallization diagram. At higher temperatures, solute and solvent molecules form elongated, prolate structures that collapse into more spherical-like aggregates upon cooling. The cooling process is accompanied by a decrease in their water content, signaled by the evolution of the solution properties, and LLPS subsequently occurs through the coalescence of the spherical-like aggregates. This spontaneously formed liquid biphasic system remains stable over a narrow temperature range without significant changes in colloidal phase composition. Further cooling culminates on a rearrangement of the molecules within the colloidal phase, leading to the crystallization of hydrate H3.

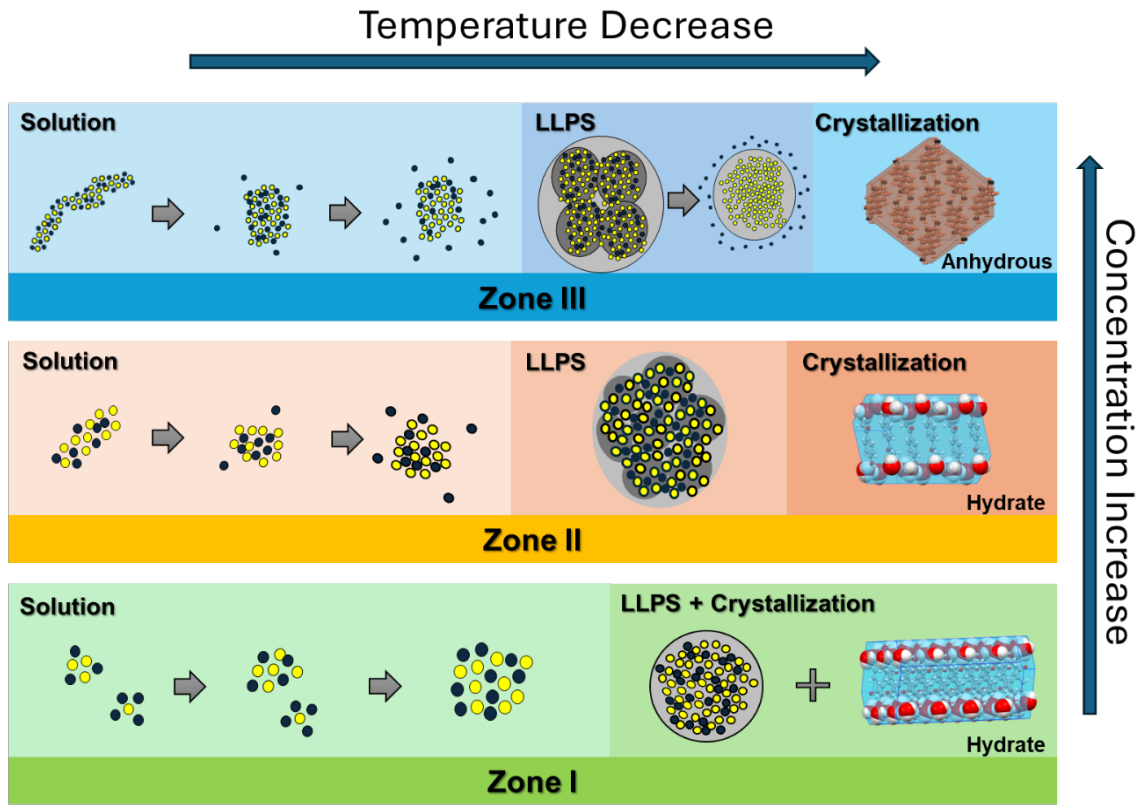

**Figure 13**. Schematic representation of the effect of concentration and temperature on the crystallization mechanism of HAP (open dots) from water (black dots).

- **Zone III ($x_{HAP}$ > 0.005):** The sequence of events described for Zone II is essentially the same in this concentration range, but with the following significant differences: (*i*) The HAP-water structures formed at high temperatures are much longer and can percolate throughout the entire system. (*ii*) Cooling leads to the formation of more spherical-like structures, accompanied by a more significant decrease in water content than in Zone II, which is detected by solution properties. (*iii*) This greater reduction in water content results in LLPS particles with less water, which continue to lose it, and ultimately result in the crystallization of the anhydrous Form I.

The present findings showcase the complex interplay of solute-solvent interactions and aggregation processes behind the cooling crystallization of HAP from water. The process is strongly influenced by the nature of the molecular aggregates formed in solution at the starting (higher) temperature and concentration range. The nature of these aggregates determines the structural transformations observed along the cooling pathway, including the change in molecular shape (elongated or spherical) and water content, the onset of LLPS, and the obtained crystal form.

**Acknowledgments.**

Centro de Química Estrutural is a Research Unit funded by Fundação para a Ciência e Tecnologia (FCT) through projects UIDB/00100/2020 (https://doi.org/10.54499/UIDB/00100/2020) and UIDP/00100/2020 (https://doi.org/10.54499/UIDP/00100/2020). Institute of Molecular Sciences is an Associate Laboratory funded by FCT through project LA/P/0056/2020 (https://doi.org/10.54499/LA/P/0056/2020). This work was further supported by the FCT project 2021.03239.CEECIND (https://doi.org/10.54499/2021.03239.CEECIND/CP1650/CT0003). R. A. Simões gratefully acknowledges the FCT grant SFRH/BPD/118771/2016.

**Supporting Information Available:**

Tables S1 to S7 with the volumetric and acoustic results. Table S8 presents the details of the simulation boxes used in the MD simulations. Figure S1 with the representation of the isentropic compressibility, as a function of temperature and concentration. Figure S2 with the representation of the molar volume, and isentropic compressibility in Zone I. Figure S3 contains the hydration numbers computed from volumetric and acoustic data. Figure S4 shows the probability of finding aggregates of HAP at different temperatures and compositions. Movies of thermomicroscopy experiments are also available.

Molecular Insights into the Crystallization of 4'-Hydroxyacetophenone from Water: Solute Aggregation, Liquid-Liquid Phase Separation and Polymorph Selection

Carlos E. S. Bernardes, Ricardo G. Simões, M. Soledade C. S. Santos, Pedro L. T. Melo, Ângela F.S. Santos, Stéphane Veesler, Manuel E. Minas da Piedade

**TOC Graphic**

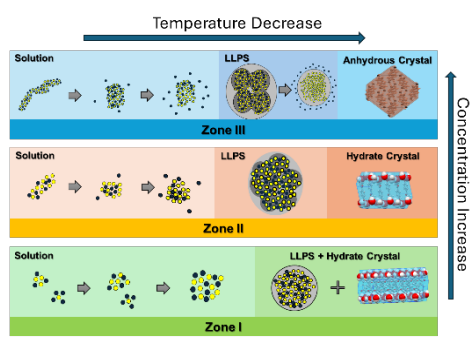

**Synopsis**

Insights into the structural rearrangements occurring in solution before the onset of the liquid-liquid phase separation (LLPS) phenomenon that mediates nucleation of different 4'-hydroxyacetophenone (HAP) polymorphs and hydrates from water, are obtained through a combination of thermomicroscopy, micro-differential scanning calorimetry, density and speed of sound measurements, and molecular dynamics simulations.

# Molecular Insights into the Crystallization of 4'-Hydroxyacetophenone from Water: Solute Aggregation, Liquid-Liquid Phase Separation and Polymorph Selection

Carlos E. S. Bernardes,[a,*] Ricardo G. Simões,[a] M. Soledade C. S. Santos,[a,*] Pedro L. T. Melo,[a] Ângela F.S. Santos,[a] Stéphane Veesler,[b] Manuel E. Minas da Piedade[a,*]

[a] *Centro de Química Estrutural, Institute of Molecular Sciences, Faculdade de Ciências, Universidade de Lisboa, Campo Grande, 1749-016 Lisboa, Portugal.*

[b] *Aix-Marseille Université, CNRS, CINaM UMR 7325, 13288 Marseille, France.*

# Supplementary Material

**Table S1.** Density, $\rho$, of aqueous solutions of 4'-hydroxyacetophenone with different molar fractions, $x_2$, as a function of temperature, $T$. Data in g·cm$^{-3}$.

| $T$ / K | $x_2$ | | | | | | | | | |
|---|---|---|---|---|---|---|---|---|---|---|
| | 0.001327 | 0.002536 | 0.003454 | 0.004126 | 0.004436 | 0.004566 | 0.004581 | 0.004743 | 0.00551 | 0.005984 | 0.006816 |
| 343.15 | 0.979035 | 0.980147 | 0.981031 | 0.981621 | 0.981892 | 0.982041 | 0.982045 | 0.982201 | 0.982919 | 0.983354 | 0.984116 |
| 341.15 | 0.980182 | 0.981304 | 0.982195 | 0.982791 | 0.983065 | 0.983215 | 0.983220 | 0.983375 | 0.984090 | 0.984540 | 0.985309 |
| 339.15 | 0.981305 | 0.982440 | 0.983338 | 0.983939 | 0.984214 | 0.984367 | 0.984369 | 0.984530 | 0.985260 | 0.985703 | 0.986476 |
| 337.15 | 0.982404 | 0.983551 | 0.984456 | 0.985064 | 0.985340 | 0.984931 | 0.985498 | 0.985660 | 0.986396 | 0.986844 | 0.987623 |
| 335.15 | 0.983483 | 0.984640 | 0.985555 | 0.986168 | 0.986445 | 0.985495 | 0.986606 | 0.986769 | 0.987511 | 0.987961 | 0.988747 |
| 333.15 | 0.984538 | 0.985707 | 0.986629 | 0.987248 | 0.987529 | 0.986601 | 0.987691 | 0.987854 | 0.988603 | 0.989060 | 0.989852 |
| 331.15 | 0.985569 | 0.986749 | 0.987682 | 0.988305 | 0.988590 | 0.987688 | 0.988751 | 0.988918 | 0.989674 | 0.990134 | 0.990931 |
| 329.15 | 0.986576 | 0.987768 | 0.988708 | 0.989337 | 0.989626 | 0.988746 | 0.989789 | 0.989956 | 0.990717 | 0.991183 | 0.991988 |
| 327.15 | 0.987559 | 0.988763 | 0.989712 | 0.990346 | 0.990639 | 0.989785 | 0.990801 | 0.990970 | 0.991739 | 0.992208 | 0.993023 |
| 325.15 | 0.988515 | 0.989733 | 0.990691 | 0.991331 | 0.991625 | 0.990797 | 0.991791 | 0.991962 | 0.992737 | 0.993211 | 0.994048* |
| 323.15 | 0.989447 | 0.990677 | 0.991646 | 0.992291 | 0.992587 | 0.991785 | 0.992754 | 0.992928 | 0.993710 | 0.994188 | 0.995075* |
| 321.15 | 0.990353 | 0.991597 | 0.992572 | 0.993227 | 0.993525 | 0.992749 | 0.993693 | 0.993868 | 0.994655* | 0.995144* | 0.996066* |
| 319.15 | 0.991234 | 0.992488 | 0.993474 | 0.994135 | 0.994435 | 0.993689 | 0.994605 | 0.994783 | 0.995577* | 0.996077* | 0.997034* |
| 317.15 | 0.992086 | 0.993354 | 0.994350 | 0.995016 | 0.995319 | 0.994601* | 0.995491* | 0.995670* | 0.996476* | 0.996981* | |
| 315.15 | 0.992911 | 0.994192 | 0.995198* | 0.995871* | 0.996176* | 0.995486* | 0.996350* | 0.996531* | 0.997355* | 0.997855* | |
| 313.15 | 0.993705 | 0.995000* | 0.996016* | 0.996696* | 0.997005* | 0.996345* | 0.997180* | 0.997363* | 0.998226* | 0.998700* | |
| 311.15 | 0.994472 | 0.995780* | 0.996807* | 0.997493* | 0.997805* | 0.997175* | 0.997980* | 0.998167* | | | |
| 309.15 | 0.995209 | 0.996531* | 0.997569* | 0.998260* | 0.998575* | 0.997975* | 0.998751* | 0.998944* | | | |
| 307.15 | 0.995913 | 0.997250* | 0.998298* | 0.998996* | 0.999316* | 0.998748* | 0.999495* | 0.999697* | | | |
| 305.15 | 0.996586 | 0.997936* | 0.998996* | 0.999700* | 1.000025* | 0.999490* | 1.000227* | 1.000417* | | | |
| 303.15 | 0.997224 | 0.998591* | 0.999660* | 1.000374* | 1.000714* | 1.000225* | 1.000934* | | | | |
| 301.15 | 0.997827 | 0.999210* | 1.000291* | 1.001011* | 1.001380* | | | | | | |
| 299.15 | 0.998396* | | | | | | | | | | |

* slight instability in $\rho$ measurements and increasing $u$ values.

**Table S2.** Speed of sound, $u$, of aqueous solutions of 4'-hydroxyacetophenone with different molar fractions, $x_2$, as a function of temperature, $T$. Data in m·s$^{-1}$.

| $T$ / K | $x_2$ | | | | | | | | | |
|---|---|---|---|---|---|---|---|---|---|---|
| | 0.001327 | 0.002536 | 0.003454 | 0.004126 | 0.004436 | 0.004566 | 0.004581 | 0.004743 | 0.00551 | 0.005984 | 0.006816 |
| 343.15 | 1554.74 | 1554.60 | 1554.79 | 1554.77 | 1554.50 | 1554.49 | 1554.49 | 1554.43 | 1554.67 | 1554.38 | 1554.52 |
| 341.15 | 1554.47 | 1554.41 | 1554.68 | 1554.71 | 1554.45 | 1554.46 | 1554.46 | 1554.44 | 1554.70 | 1554.46 | 1554.64 |
| 339.15 | 1554.04 | 1554.08 | 1554.41 | 1554.49 | 1554.23 | 1554.28 | 1554.26 | 1554.28 | 1554.58 | 1554.38 | 1554.60 |
| 337.15 | 1553.42 | 1553.58 | 1553.97 | 1554.10 | 1553.84 | 1554.12 | 1553.89 | 1553.94 | 1554.28 | 1554.12 | 1554.39 |
| 335.15 | 1552.63 | 1552.89 | 1553.35 | 1553.53 | 1553.30 | 1553.92 | 1553.35 | 1553.44 | 1553.81 | 1553.69 | 1554.01 |
| 333.15 | 1551.65 | 1552.03 | 1552.57 | 1552.79 | 1552.59 | 1553.39 | 1552.65 | 1552.76 | 1553.17 | 1553.09 | 1553.46 |
| 331.15 | 1550.50 | 1550.96 | 1551.59 | 1551.87 | 1551.69 | 1552.68 | 1551.75 | 1551.90 | 1552.35 | 1552.31 | 1552.73 |
| 329.15 | 1549.17 | 1549.74 | 1550.43 | 1550.76 | 1550.60 | 1551.80 | 1550.68 | 1550.85 | 1551.35 | 1551.34 | 1551.82 |
| 327.15 | 1547.63 | 1548.32 | 1549.09 | 1549.47 | 1549.32 | 1550.72 | 1549.41 | 1549.61 | 1550.16 | 1550.18 | 1550.72 |
| 325.15 | 1545.90 | 1546.71 | 1547.55 | 1547.98 | 1547.91 | 1549.46 | 1547.95 | 1548.18 | 1548.78 | 1548.83 | 1549.37* |
| 323.15 | 1543.96 | 1544.91 | 1545.82 | 1546.30 | 1546.59 | 1548.00 | 1546.30 | 1546.69 | 1547.21 | 1547.30 | 1547.78* |
| 321.15 | 1541.83 | 1542.94 | 1543.87 | 1544.40 | 1544.96 | 1546.35 | 1544.45 | 1545.13 | 1545.43* | 1545.55* | 1545.98* |
| 319.15 | 1539.47 | 1540.76 | 1541.73 | 1542.32 | 1543.35 | 1544.50 | 1542.43 | 1543.47 | 1543.45* | 1543.41* | 1544.03* |
| 317.15 | 1536.90 | 1538.38 | 1539.39 | 1540.28 | 1541.25 | 1542.45* | 1540.24* | 1541.36* | 1541.23* | 1541.30* | |
| 315.15 | 1534.11 | 1535.74 | 1536.85 | 1537.49* | 1538.85 | 1540.28* | 1537.84* | 1538.98* | 1538.73* | 1538.82* | |
| 313.15 | 1531.09 | 1532.86* | 1534.09* | 1534.78* | 1536.22* | 1537.86* | 1535.20* | 1536.36* | 1536.11* | 1536.06* | |
| 311.15 | 1527.84 | 1529.75* | 1531.09* | 1531.91* | 1533.37* | 1536.23* | 1532.39* | 1533.51* | | | |
| 309.15 | 1524.34 | 1526.40* | 1527.88* | 1528.81* | 1530.27* | 1532.36* | 1529.33* | 1530.43* | | | |
| 307.15 | 1520.60 | 1522.80* | 1524.43* | 1525.48* | 1526.93* | 1529.44* | 1525.98* | 1527.00* | | | |
| 305.15 | 1516.59 | 1518.95* | 1520.69* | 1521.89* | 1523.35* | 1526.08* | 1522.26* | 1523.30* | | | |
| 303.15 | 1512.32 | 1514.84* | 1516.70* | 1518.01* | 1519.41* | 1522.39* | 1518.01* | | | | |
| 301.15 | 1507.79 | 1510.47* | 1512.43* | 1513.81* | 1512.10* | | | | | | |
| 299.15 | 1502.99* | | | | | | | | | | |

* slight instability in $\rho$ measurements and increasing $u$ values.

**Table S3.** Molar volume, $V_m$, of aqueous solutions of 4'-hydroxyacetophenone with different molar fractions, $x_2$, as a function of temperature, $T$. Data in cm³·mol⁻¹.

| T / K | $x_2$ | | | | | | | | | | |
|---|---|---|---|---|---|---|---|---|---|---|---|
| | 0.001327 | 0.002536 | 0.003454 | 0.004126 | 0.004436 | 0.004566 | 0.004581 | 0.004743 | 0.00551 | 0.005984 | 0.006816 |
| 343.15 | 18.56090 | 18.68550 | 18.77929 | 18.84888 | 18.88092 | 18.89367 | 18.89542 | 18.91189 | 18.99032 | 19.03887 | 19.12398 |
| 341.15 | 18.53918 | 18.66347 | 18.75703 | 18.82644 | 18.85839 | 18.87111 | 18.87284 | 18.88931 | 18.96772 | 19.01594 | 19.10082 |
| 339.15 | 18.51796 | 18.64189 | 18.73523 | 18.80448 | 18.83638 | 18.84903 | 18.85081 | 18.86715 | 18.94520 | 18.99350 | 19.07823 |
| 337.15 | 18.49725 | 18.62083 | 18.71395 | 18.78300 | 18.81485 | 18.83823 | 18.82922 | 18.84552 | 18.92338 | 18.97154 | 19.05607 |
| 335.15 | 18.47695 | 18.60023 | 18.69309 | 18.76198 | 18.79378 | 18.82745 | 18.80807 | 18.82434 | 18.90201 | 18.95009 | 19.03441 |
| 333.15 | 18.45716 | 18.58010 | 18.67274 | 18.74145 | 18.77315 | 18.80635 | 18.78741 | 18.80366 | 18.88114 | 18.92904 | 19.01316 |
| 331.15 | 18.43785 | 18.56048 | 18.65283 | 18.72141 | 18.75300 | 18.78565 | 18.76727 | 18.78343 | 18.86070 | 18.90851 | 18.99245 |
| 329.15 | 18.41903 | 18.54133 | 18.63347 | 18.70188 | 18.73337 | 18.76555 | 18.74759 | 18.76374 | 18.84085 | 18.88849 | 18.97222 |
| 327.15 | 18.40069 | 18.52267 | 18.61457 | 18.68282 | 18.71421 | 18.74585 | 18.72844 | 18.74454 | 18.82143 | 18.86899 | 18.95244 |
| 325.15 | 18.38290 | 18.50452 | 18.59618 | 18.66426 | 18.69560 | 18.72670 | 18.70974 | 18.72579 | 18.80251 | 18.84993 | 18.93290 |
| 323.15 | 18.36558 | 18.48689 | 18.57827 | 18.64620 | 18.67748 | 18.70805 | 18.6916 | 18.70757 | 18.78410 | 18.83140 | 18.91336 |
| 321.15 | 18.34878 | 18.46973 | 18.56093 | 18.62863 | 18.65985 | 18.68988 | 18.67393 | 18.68988 | 18.76625 | 18.81331 | 18.89454 |
| 319.15 | 18.33247 | 18.45315 | 18.54408 | 18.61162 | 18.64277 | 18.67220 | 18.65681 | 18.67269 | 18.74887 | 18.79569 | 18.87620 |
| 317.15 | 18.31673 | 18.43707 | 18.52775 | 18.59514 | 18.62622 | 18.65508 | 18.64021 | 18.65605 | 18.73196 | 18.77865 | |
| 315.15 | 18.30151 | 18.42153 | 18.51196 | 18.57917 | 18.61019 | 18.63849 | 18.62413 | 18.63994 | 18.71545 | 18.76220 | |
| 313.15 | 18.28689 | 18.40657 | 18.49676 | 18.56379 | 18.59472 | 18.62242 | 18.60863 | 18.62439 | 18.69912 | 18.74632 | |
| 311.15 | 18.27278 | 18.39215 | 18.48208 | 18.54896 | 18.57981 | 18.60692 | 18.59372 | 18.60938 | | | |
| 309.15 | 18.25925 | 18.37829 | 18.46796 | 18.53471 | 18.56548 | 18.59201 | 18.57936 | 18.59491 | | | |
| 307.15 | 18.24634 | 18.36504 | 18.45447 | 18.52105 | 18.55172 | 18.57762 | 18.56553 | 18.58090 | | | |
| 305.15 | 18.23402 | 18.35241 | 18.44158 | 18.50801 | 18.53856 | 18.56383 | 18.55195 | 18.56753 | | | |
| 303.15 | 18.22236 | 18.34038 | 18.42933 | 18.49554 | 18.52580 | 18.55019 | 18.53884 | | | | |
| 301.15 | 18.21134 | 18.32901 | 18.41770 | 18.48377 | 18.51348 | | | | | | |
| 299.15 | 18.20097 | | | | | | | | | | |

**Table S4.** Isentropic compressibility, $K_S$, of aqueous solutions of 4'-hydroxyacetophenone with different molar fractions, $x_2$, as a function of temperature, $T$. Data in $10^{10}$ Pa⁻¹.

| T / K | $x_2$ | | | | | | | | | | |
|---|---|---|---|---|---|---|---|---|---|---|---|
| | 0.001327 | 0.002536 | 0.003454 | 0.004126 | 0.004436 | 0.004566 | 0.004581 | 0.004743 | 0.00551 | 0.005984 | 0.006816 |
| 343.15 | 4.2256 | 4.2215 | 4.2167 | 4.2143 | 4.2146 | 4.2140 | 4.2140 | 4.2136 | 4.2093 | 4.2090 | 4.2049 |
| 341.15 | 4.2221 | 4.2176 | 4.2123 | 4.2096 | 4.2098 | 4.2091 | 4.2091 | 4.2086 | 4.2041 | 4.2035 | 4.1992 |
| 339.15 | 4.2196 | 4.2145 | 4.2089 | 4.2059 | 4.2061 | 4.2052 | 4.2053 | 4.2045 | 4.1998 | 4.1990 | 4.1945 |
| 337.15 | 4.2182 | 4.2125 | 4.2065 | 4.2032 | 4.2034 | 4.2036 | 4.2025 | 4.2015 | 4.1966 | 4.1955 | 4.1907 |
| 335.15 | 4.2179 | 4.2115 | 4.2051 | 4.2015 | 4.2016 | 4.2023 | 4.2006 | 4.1995 | 4.1943 | 4.1930 | 4.1880 |
| 333.15 | 4.2187 | 4.2117 | 4.2048 | 4.2009 | 4.2009 | 4.2005 | 4.1999 | 4.1985 | 4.1932 | 4.1917 | 4.1863 |
| 331.15 | 4.2206 | 4.2130 | 4.2056 | 4.2015 | 4.2012 | 4.1997 | 4.2002 | 4.1987 | 4.1930 | 4.1913 | 4.1856 |
| 329.15 | 4.2235 | 4.2153 | 4.2075 | 4.2031 | 4.2027 | 4.2000 | 4.2016 | 4.2000 | 4.1940 | 4.1921 | 4.1861 |
| 327.15 | 4.2277 | 4.2188 | 4.2105 | 4.2058 | 4.2053 | 4.2014 | 4.2042 | 4.2024 | 4.1962 | 4.1941 | 4.1877 |
| 325.15 | 4.2331 | 4.2234 | 4.2147 | 4.2097 | 4.2088 | 4.2039 | 4.2079 | 4.2059 | 4.1994 | 4.1971 | 4.1907 |
| 323.15 | 4.2397 | 4.2292 | 4.2201 | 4.2148 | 4.2119 | 4.2077 | 4.2128 | 4.2100 | 4.2038 | 4.2013 | 4.1950 |
| 321.15 | 4.2475 | 4.2361 | 4.2269 | 4.2211 | 4.2168 | 4.2125 | 4.2189 | 4.2145 | 4.2095 | 4.2068 | 4.2005 |
| 319.15 | 4.2568 | 4.2443 | 4.2347 | 4.2287 | 4.2218 | 4.2187 | 4.2261 | 4.2197 | 4.2164 | 4.2145 | 4.2070 |
| 317.15 | 4.2673 | 4.2537 | 4.2439 | 4.2361 | 4.2295 | 4.2260 | 4.2343 | 4.2274 | 4.2247 | 4.2222 | |
| 315.15 | 4.2793 | 4.2647 | 4.2543 | 4.2479 | 4.2391 | 4.2341 | 4.2439 | 4.2369 | 4.2348 | 4.2321 | |
| 313.15 | 4.2928 | 4.2773 | 4.2661 | 4.2594 | 4.2501 | 4.2438 | 4.2550 | 4.2478 | 4.2455 | 4.2438 | |
| 311.15 | 4.3078 | 4.2913 | 4.2795 | 4.2720 | 4.2625 | 4.2493 | 4.2672 | 4.2601 | | | |
| 309.15 | 4.3243 | 4.3070 | 4.2942 | 4.2860 | 4.2764 | 4.2674 | 4.2810 | 4.2740 | | | |
| 307.15 | 4.3426 | 4.3242 | 4.3105 | 4.3015 | 4.2920 | 4.2804 | 4.2966 | 4.2900 | | | |
| 305.15 | 4.3626 | 4.3432 | 4.3287 | 4.3188 | 4.3091 | 4.2960 | 4.3144 | 4.3077 | | | |
| 303.15 | 4.3845 | 4.3639 | 4.3486 | 4.3380 | 4.3285 | 4.3137 | 4.3356 | | | | |
| 301.15 | 4.4082 | 4.3865 | 4.3704 | 4.3593 | 4.3676 | | | | | | |
| 299.15 | 4.4339 | | | | | | | | | | |

**Table S5.** Apparent molar volume, $V_{\phi,2}$, of aqueous solutions of 4'-hydroxyacetophenone with different molar fractions, $x_2$, as a function of temperature, $T$. Data in $10^4$ m$^3$·mol$^{-1}$.

| $T$ / K | $x_2$ | | | | | | | | | |
|---|---|---|---|---|---|---|---|---|---|---|
| | 0.001327 | 0.002536 | 0.003454 | 0.004126 | 0.004436 | 0.004566 | 0.004581 | 0.004743 | 0.00551 | 0.005984 | 0.006816 |
| 343.15 | 1.2099 | 1.2125 | 1.2105 | 1.2120 | 1.2125 | 1.2112 | 1.2117 | 1.2113 | 1.2106 | 1.2104 | 1.2101 |
| 341.15 | 1.2065 | 1.2093 | 1.2075 | 1.2090 | 1.2095 | 1.2082 | 1.2086 | 1.2084 | 1.2080 | 1.2075 | 1.2071 |
| 339.15 | 1.2035 | 1.2062 | 1.2045 | 1.2061 | 1.2067 | 1.2053 | 1.2058 | 1.2054 | 1.2047 | 1.2046 | 1.2043 |
| 337.15 | 1.2005 | 1.2032 | 1.2017 | 1.2032 | 1.2038 | 1.2040 | 1.2029 | 1.2025 | 1.2019 | 1.2017 | 1.2015 |
| 335.15 | 1.1973 | 1.2003 | 1.1986 | 1.2002 | 1.2010 | 1.2024 | 1.2000 | 1.1996 | 1.1990 | 1.1989 | 1.1987 |
| 333.15 | 1.1942 | 1.1973 | 1.1957 | 1.1974 | 1.1980 | 1.1996 | 1.1971 | 1.1967 | 1.1962 | 1.1960 | 1.1958 |
| 331.15 | 1.1912 | 1.1944 | 1.1927 | 1.1945 | 1.1951 | 1.1966 | 1.1942 | 1.1938 | 1.1932 | 1.1931 | 1.1930 |
| 329.15 | 1.1882 | 1.1914 | 1.1899 | 1.1917 | 1.1923 | 1.1938 | 1.1914 | 1.1910 | 1.1905 | 1.1903 | 1.1902 |
| 327.15 | 1.1853 | 1.1885 | 1.1870 | 1.1889 | 1.1894 | 1.1909 | 1.1886 | 1.1882 | 1.1877 | 1.1876 | 1.1874 |
| 325.15 | 1.1825 | 1.1856 | 1.1841 | 1.1860 | 1.1866 | 1.1881 | 1.1857 | 1.1853 | 1.1849 | 1.1847 | 1.1841 |
| 323.15 | 1.1795 | 1.1827 | 1.1812 | 1.1831 | 1.1838 | 1.1853 | 1.1829 | 1.1825 | 1.1820 | 1.1819 | 1.1801 |
| 321.15 | 1.1767 | 1.1797 | 1.1784 | 1.1802 | 1.1810 | 1.1825 | 1.1801 | 1.1797 | 1.1793 | 1.1790 | 1.1765 |
| 319.15 | 1.1735 | 1.1769 | 1.1755 | 1.1774 | 1.1781 | 1.1796 | 1.1773 | 1.1768 | 1.1765 | 1.1759 | 1.1728 |
| 317.15 | 1.1705 | 1.1739 | 1.1726 | 1.1745 | 1.1753 | 1.1768 | 1.1744 | 1.1740 | 1.1735 | 1.1730 | |
| 315.15 | 1.1675 | 1.1710 | 1.1697 | 1.1717 | 1.1725 | 1.1740 | 1.1716 | 1.1711 | 1.1704 | 1.1701 | |
| 313.15 | 1.1648 | 1.1682 | 1.1669 | 1.1689 | 1.1697 | 1.1711 | 1.1688 | 1.1683 | 1.1665 | 1.1673 | |
| 311.15 | 1.1618 | 1.1653 | 1.1640 | 1.1660 | 1.1669 | 1.1683 | 1.1660 | 1.1655 | | | |
| 309.15 | 1.1586 | 1.1623 | 1.1610 | 1.1632 | 1.1640 | 1.1656 | 1.1632 | 1.1625 | | | |
| 307.15 | 1.1557 | 1.1594 | 1.1581 | 1.1603 | 1.1611 | 1.1627 | 1.1603 | 1.1593 | | | |
| 305.15 | 1.1526 | 1.1565 | 1.1552 | 1.1575 | 1.1583 | 1.1598 | 1.1565 | 1.1561 | | | |
| 303.15 | 1.1497 | 1.1534 | 1.1523 | 1.1546 | 1.1548 | 1.1559 | 1.1525 | | | | |
| 301.15 | 1.1469 | 1.1506 | 1.1494 | 1.1518 | 1.1510 | | | | | | |
| 299.15 | 1.1438 | | | | | | | | | | |

**Table S6.** Relative association, $RA$, of aqueous solutions of 4'-hydroxyacetophenone with different molar fractions, $x_2$, as a function of temperature, $T$.

| $T$ / K | $x_2$ | | | | | | | | | |
|---|---|---|---|---|---|---|---|---|---|---|
| | 0.001327 | 0.002536 | 0.003454 | 0.004126 | 0.004436 | 0.004566 | 0.004581 | 0.004743 | 0.00551 | 0.005984 | 0.006816 |
| 343.15 | 1.001387 | 1.002554 | 1.003418 | 1.004025 | 1.004360 | 1.004515 | 1.004520 | 1.004692 | 1.005375 | 1.005883 | 1.006631 |
| 341.15 | 1.001369 | 1.002529 | 1.003380 | 1.003982 | 1.004320 | 1.004469 | 1.004476 | 1.004639 | 1.005312 | 1.005823 | 1.006570 |
| 339.15 | 1.001272 | 1.002421 | 1.003266 | 1.003862 | 1.004274 | 1.004420 | 1.004426 | 1.004511 | 1.005192 | 1.005687 | 1.006427 |
| 337.15 | 1.001328 | 1.002463 | 1.003301 | 1.003894 | 1.004230 | 1.004395 | 1.004381 | 1.004534 | 1.005213 | 1.005703 | 1.006439 |
| 335.15 | 1.001313 | 1.002433 | 1.003266 | 1.003851 | 1.004183 | 1.004371 | 1.004336 | 1.004482 | 1.005159 | 1.005642 | 1.006374 |
| 333.15 | 1.001294 | 1.002403 | 1.003223 | 1.003804 | 1.004135 | 1.004323 | 1.004287 | 1.004428 | 1.005101 | 1.005583 | 1.006309 |
| 331.15 | 1.001275 | 1.002375 | 1.003186 | 1.003760 | 1.004088 | 1.004276 | 1.004238 | 1.004376 | 1.005046 | 1.005522 | 1.006240 |
| 329.15 | 1.001257 | 1.002343 | 1.003147 | 1.003714 | 1.004043 | 1.004223 | 1.004191 | 1.004324 | 1.004987 | 1.005463 | 1.006174 |
| 327.15 | 1.001236 | 1.002308 | 1.003104 | 1.003664 | 1.003993 | 1.004178 | 1.004139 | 1.004266 | 1.004927 | 1.005397 | 1.006106 |
| 325.15 | 1.001216 | 1.002274 | 1.003062 | 1.003617 | 1.003930 | 1.004124 | 1.004090 | 1.004212 | 1.004868 | 1.005337 | 1.006067 |
| 323.15 | 1.001198 | 1.002238 | 1.003022 | 1.003570 | 1.003806 | 1.004073 | 1.004038 | 1.004130 | 1.004808 | 1.005273 | 1.006065 |
| 321.15 | 1.001175 | 1.002191 | 1.002975 | 1.003522 | 1.003702 | 1.004022 | 1.003983 | 1.004012 | 1.004742 | 1.005210 | 1.006047 |
| 319.15 | 1.001157 | 1.002144 | 1.002929 | 1.003469 | 1.003548 | 1.003968 | 1.003919 | 1.003874 | 1.004679 | 1.005192 | 1.006022 |
| 317.15 | 1.001135 | 1.002095 | 1.002878 | 1.003358 | 1.003453 | 1.003911 | 1.003845 | 1.003783 | 1.004623 | 1.005118 | |
| 315.15 | 1.001113 | 1.002049 | 1.002821 | 1.003362 | 1.003372 | 1.003832 | 1.003768 | 1.003703 | 1.004587 | 1.005070 | |
| 313.15 | 1.001088 | 1.002006 | 1.002763 | 1.003297 | 1.003293 | 1.003759 | 1.003692 | 1.003623 | 1.004546 | 1.005035 | |
| 311.15 | 1.001062 | 1.001960 | 1.002703 | 1.003214 | 1.003208 | 1.003463 | 1.003599 | 1.003541 | | | |
| 309.15 | 1.001037 | 1.001917 | 1.002637 | 1.003127 | 1.003124 | 1.003599 | 1.003507 | 1.003462 | | | |
| 307.15 | 1.001009 | 1.001869 | 1.002564 | 1.003035 | 1.003038 | 1.003481 | 1.003427 | 1.003405 | | | |
| 305.15 | 1.000982 | 1.001818 | 1.002499 | 1.002942 | 1.002948 | 1.003399 | 1.003389 | 1.003352 | | | |
| 303.15 | 1.000949 | 1.001766 | 1.002428 | 1.002856 | 1.002888 | 1.003360 | 1.003417 | | | | |
| 301.15 | 1.000913 | 1.001706 | 1.002356 | 1.002773 | 1.003521 | | | | | | |
| 299.15 | 1.000874 | | | | | | | | | | |

**Table S7.** Acoustic impedance, $Z$, of aqueous solutions of 4'-hydroxyacetophenone with different molar fractions, $x_2$, as a function of temperature, $T$. Data in ·10$^{-6}$ Pa·s·m$^{-3}$.

| $T$ / K | $x_2$ | | | | | | | | | | |
|---|---|---|---|---|---|---|---|---|---|---|---|
| | 0.001327 | 0.002536 | 0.003454 | 0.004126 | 0.004436 | 0.004566 | 0.004581 | 0.004743 | 0.00551 | 0.005984 | 0.006816 |
| 343.15 | 1.522142 | 1.523736 | 1.525293 | 1.526197 | 1.526351 | 1.526572 | 1.526576 | 1.526761 | 1.528113 | 1.528501 | 1.529830 |
| 341.15 | 1.523666 | 1.525344 | 1.527003 | 1.527959 | 1.528122 | 1.528371 | 1.528373 | 1.528592 | 1.529966 | 1.530429 | 1.531804 |
| 339.15 | 1.524983 | 1.526792 | 1.528513 | 1.529523 | 1.529697 | 1.529978 | 1.529963 | 1.530231 | 1.531660 | 1.532152 | 1.533574 |
| 337.15 | 1.526086 | 1.528023 | 1.529816 | 1.530883 | 1.531062 | 1.530699 | 1.531355 | 1.531661 | 1.533131 | 1.533675 | 1.535150 |
| 335.15 | 1.526983 | 1.529042 | 1.530917 | 1.532045 | 1.532248 | 1.531381 | 1.532549 | 1.532891 | 1.534403 | 1.534989 | 1.536522 |
| 333.15 | 1.527662 | 1.529842 | 1.531809 | 1.532992 | 1.533223 | 1.532576 | 1.533533 | 1.533899 | 1.535466 | 1.536097 | 1.537693 |
| 331.15 | 1.528124 | 1.530408 | 1.532480 | 1.533719 | 1.533982 | 1.533563 | 1.534297 | 1.534697 | 1.536320 | 1.536989 | 1.538652 |
| 329.15 | 1.528369 | 1.530780 | 1.532923 | 1.534225 | 1.534510 | 1.534331 | 1.534841 | 1.535268 | 1.536949 | 1.537658 | 1.539391 |
| 327.15 | 1.528374 | 1.530919 | 1.533151 | 1.534511 | 1.534819 | 1.534879 | 1.535153 | 1.535615 | 1.537351 | 1.538098 | 1.539905 |
| 325.15 | 1.528142 | 1.530830 | 1.533146 | 1.534562 | 1.534945 | 1.535198 | 1.535241 | 1.535740 | 1.537532 | 1.538316 | 1.540146 |
| 323.15 | 1.527669 | 1.530505 | 1.532903 | 1.534378 | 1.535127 | 1.535281 | 1.535096 | 1.535749 | 1.537475 | 1.538302 | 1.540153 |
| 321.15 | 1.526953 | 1.529978 | 1.532401 | 1.533943 | 1.534957 | 1.535139 | 1.534707 | 1.535654 | 1.537169 | 1.538044 | 1.539900 |
| 319.15 | 1.525977 | 1.529184 | 1.531669 | 1.533271 | 1.534760 | 1.534749 | 1.534109 | 1.535414 | 1.536620 | 1.537354 | 1.539452 |
| 317.15 | 1.524741 | 1.528151 | 1.530697 | 1.532602 | 1.534031 | 1.534120 | 1.533295 | 1.534683 | 1.535799 | 1.536645 | |
| 315.15 | 1.523235 | 1.526823 | 1.529475 | 1.531137 | 1.532969 | 1.533326 | 1.532226 | 1.533638 | 1.534655 | 1.535522 | |
| 313.15 | 1.521453 | 1.525198 | 1.527974 | 1.529706 | 1.531622 | 1.532236 | 1.530870 | 1.532309 | 1.533384 | 1.534060 | |
| 311.15 | 1.519391 | 1.523299 | 1.526196 | 1.528065 | 1.530003 | 1.531889 | 1.529290 | 1.530700 | | | |
| 309.15 | 1.517041 | 1.521102 | 1.524164 | 1.526151 | 1.528091 | 1.529257 | 1.527420 | 1.528809 | | | |
| 307.15 | 1.514380 | 1.518612 | 1.521835 | 1.523946 | 1.525887 | 1.527523 | 1.525208 | 1.526539 | | | |
| 305.15 | 1.511411 | 1.515815 | 1.519164 | 1.521437 | 1.523389 | 1.525301 | 1.522610 | 1.523935 | | | |
| 303.15 | 1.508125 | 1.512704 | 1.516182 | 1.518574 | 1.520496 | 1.522728 | 1.519426 | | | | |
| 301.15 | 1.504508 | 1.509275 | 1.512871 | 1.515342 | 1.514186 | | | | | | |
| 299.15 | 1.500580 | | | | | | | | | | |

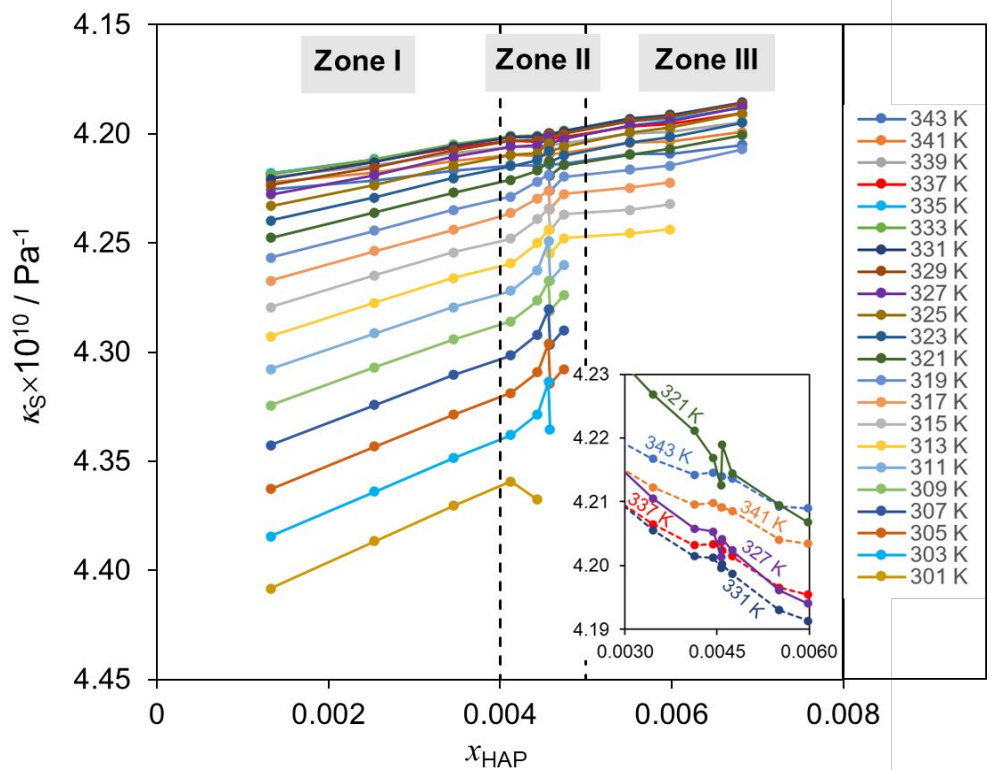

**Figure S1.** Isentropic compressibility, $\kappa_s$, as a function of temperature and HAP molar fraction, $x_{HAP}$, in water. Solid lines and filled markers refer to measurements where $\kappa_s$ decreases with temperature. Dash lines in the inset indicate measurements where $\kappa_s$ increases with temperature.

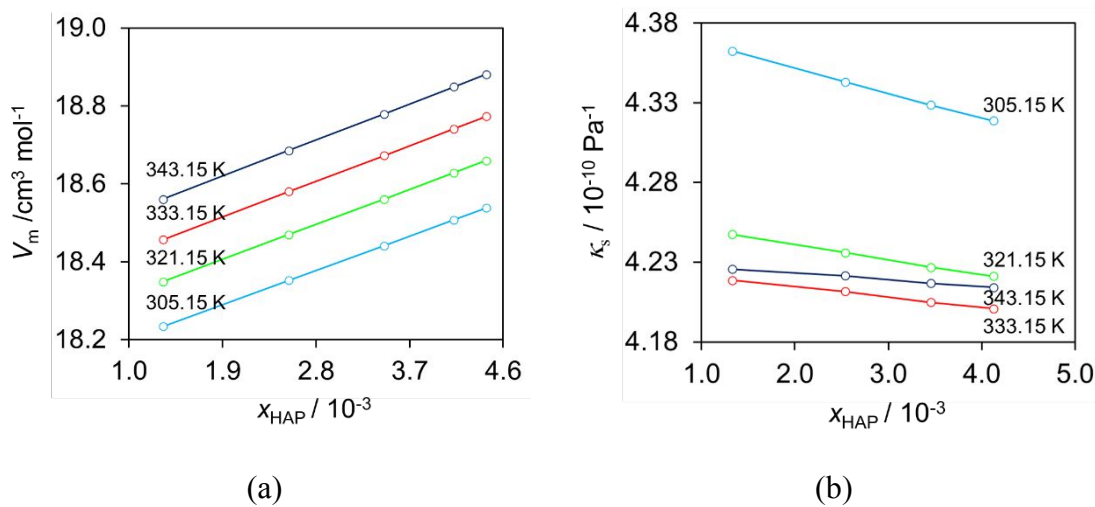

(a)  (b)

**Figure S2.** Dependence of (a) the molar volume, $V_m$, and (b) isentropic compressibility, $\kappa_s$, for mole fractions of HAP, $x_{HAP}$, in zone I of concentrations and at different temperatures.

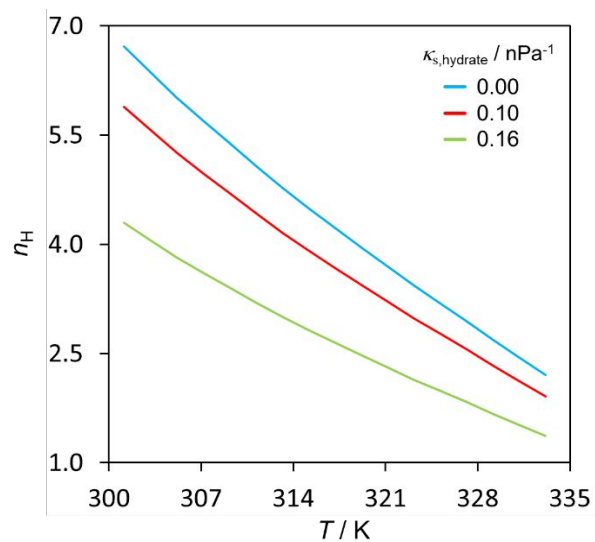

**Figure S3.** Dependence of the hydration number, $n_H$, as a function of temperature for $x_{HAP} = 0.001327$. The different lines were obtained considering different values of $\kappa_{s,hydrate}$: zero (green), $1.2 \cdot 10^{-10}$ Pa$^{-1}$ (red), $1.60 \cdot 10^{-10}$ Pa$^{-1}$ (blue).[1-2]

**Table S8.** Details of the simulation boxes used in the molecular dynamics studies: $x_{HAP}$ is the molar fraction of 4'-hydroxyacetophenone (HAP), $N_{HAP}$ and $N_W$ the number of HAP and water molecules in the simulation box, and $l_{box}$ is the average side length of the cubic box after equilibration of the system at 360 K.

| $x_{HAP}$ | $N_{HAP}$ | $N_W$ | $l_{box}$ / Å |
|---|---|---|---|
| 0.006 | 300 | 49700 | 117.502 |
| 0.004 | 200 | 49800 | 117.232 |
| 0.002 | 100 | 49900 | 116.990 |

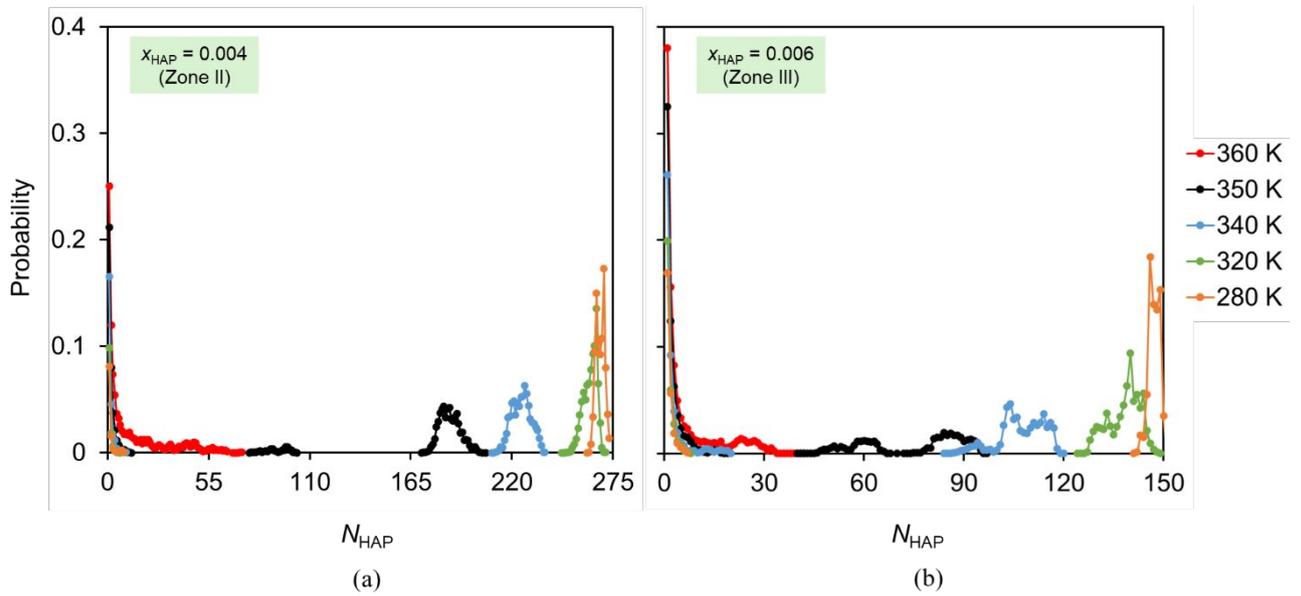

**Figure S4.** Probability of finding aggregates containing $N_{HAP}$ molecules of HAP, at different temperatures and compositions: (a) $x_{HAP}$ = 0.004 and (b) $x_{HAP}$ = 0.006. The simulation boxes contained 200 and 300 molecules of solute for $x_{HAP}$ = 0.004 and $x_{HAP}$ = 0.006, respectively.